\documentstyle[12pt,ssi]{article}

\def\etal{{\it et al.}}
\def\ie{{\it i.e.}}
\def\eg{{\it e.g.}}
\def\gessim{\mathrel {\vcenter {\baselineskip 0pt \kern 0pt
\hbox{$>$} \kern 0pt \hbox{$\sim$} }}}

\def\mttbar {t\, \overline{t}}

\def\ttbar {$\mttbar$}

\def\Et {$E_T$}
\def\mMEt{\not\kern-.35em {E_T}}
\def\MEt{\hbox{$\mMEt$}}
\def\Pt {$P_T$}
\def\invpb {{pb$^{-1}$}}
\def\invfb {{fb$^{-1}$}}
\def\GeVc  {$\rm GeV\!/c$}

\def\TeVcc {$\rm TeV\!/c^2$}
\def\GeVccm {\rm GeV\!/c^2}
\def\GeVcc {$\GeVccm$}

\def \rightdownarrow
 {\kern.4em {\raise1.75ex\hbox{$\Bigl |$}$\kern-0.27em{\longrightarrow}$}}
\def \mrightdownarrow
 {\kern.4em {\raise1.25ex \hbox{$|$} \kern-0.27em{
                                                \longrightarrow}}}

\def\DZero{D$\mbox{0\kern-.55em /}$}

\def\aplanarity{{\cal A}}
\def\invpb{${\rm pb^{-1}}$}
\def\mGeVcc{{\rm GeV/c^2}}
\def\mBR{{\cal B}}

\def\invcms{${\rm cm^{-2}s^{-1}}$}
\def\pbarp{$p\bar{p}$}
\def \mrightdownarrowlong
 {\kern.4em {\vbox{\raise1.25ex \hbox{$|$}
                   \raise2.50ex \hbox{$|$}} \kern-0.27em{
                                                \longrightarrow}}}
\begin{document}

\title{
{\normalsize
\hfill CDF/DOC/TOP/PUBLIC/3801}  \\[0.1in]
TOP QUARK STUDIES AT HADRON COLLIDERS\footnote{
{\normalsize
Lectures presented at the 1995 SLAC Summer Institute}}
}

\author{Pekka K. Sinervo\thanks{Supported by the Natural Sciences and 
Engineering Council of Canada.}\\
Department of Physics \\
University of Toronto, Toronto, Ontario, Canada  M5S 1A7
}

\maketitle

\begin{abstract}
The techniques used to study top quarks at hadron colliders
are presented.
The analyses that discovered the top quark are described,
with emphasis on the techniques used to tag $b$~quark jets in candidate
events.
The most recent measurements of top quark properties by the 
CDF and \DZero\ collaborations are reviewed, including
the top quark cross section, mass, branching fractions and production properties.

Future top quark studies at hadron colliders are discussed, and 
predictions for event yields and uncertainties in the measurements of
top quark properties are presented.

\end{abstract}

\section{Introduction}
\subsection{The Case for Top}
The top quark and the Higgs boson are the heaviest 
elementary particles predicted by the standard model. \cite{ref: sm} 
The four lightest quark flavours, the up, down,
strange and charm quarks, were well-established by the mid-1970's. 
The discovery in 1977 
\cite{ref: b quark discovery}\ 
of the $\Upsilon$\ resonances, a new family of massive
hadrons, required the
introduction of the fifth quark flavour.  Experimental and
theoretical studies have indicated that this quark has a heavier
partner, the top quark.

Indirect evidence for the top quark comes from a number of sources. 
The most compelling data come from the observed properties of
the scattering process $e^+e^-\rightarrow b\bar{b}$, where the
asymmetry in the scattering of the $b$~quark relative to the incoming
electron direction implies that the $b$~quark has weak isospin 
of 0.5.  The most precise measurement of this comes from the
LEP collider, where this asymmetry has been 
found\cite{ref: LEP b quark asymmetry}
to be in excellent agreement with the standard model 
expectation of 0.100 assuming that the $b$~quark is a
member of an $SU(2)$\ doublet.  The other member of that doublet
would by definition be the top quark.

Additional indirect evidence comes from the study of $b$~quark
decays.  It has been experimentally determined that the $b$~quark
does not decay via processes that yield zero net flavour in the final
state (\eg, $b\rightarrow \mu^+\mu^-X$), or where the decay results in
only a quark of the same charge (\eg, $b\rightarrow s X$\ where $X$\
is a state with no net flavour quantum numbers) 
\cite{ref: B->mumu limits}.  The absence of
these ``flavour-changing neutral currents'' 
in the standard model implies that the $b$~quark is a member of an
$SU(2)$\ doublet.

Finally, evidence for the existence of a massive
fermion that couples via the electroweak force to the $b$~quark
comes from detailed measurements of the $Z^\circ$\ and $W^+$\ bosons
performed at LEP, SLC, the CERN S$p\bar{p}$S\ and the Fermilab Tevatron
Collider.  This body of data, and in particular the radiative mass
shifts of the electroweak bosons, can only be described in the
standard model by introducing a top quark.  A recent
compilation of data \cite{ref: recent LEP t mass}\ indicates that the
standard model top quark has a mass of
\begin{eqnarray}
M_{top} = 169^{+16}_{-18}\phantom{1}^{+17}_{-20}\ \mGeVcc.
\end{eqnarray}
The second uncertainty corresponds to variations of the unknown Higgs
boson mass between 60 and 1000 \GeVcc\ (its nominal value is 300 \GeVcc).

Taken together, these observations make a strong case for the top
quark's existence.  They also imply that our understanding of nature
via the standard model would be profoundly shaken if the top quark
was shown not to exist with its expected properties.  The
observation of the top quark is therefore of considerable
significance.

\subsection{Earlier Top Quark Searches}
Direct searches for the top quark have been performed at virtually all
of the high-energy collider facilities that have operated in the
last twenty years\cite{ref: top searches}.  The most
model-independent searches have taken place at $e^+e^-$\ colliders,
where one looks for the production and decay of a pair of massive
fermions.  Because of the relatively large mass of the top quark, its
decay yields events that are quite spherical and are relatively easy
to separate from the background of lighter quark production.  The
most stringent limits have been set by the LEP collaborations, which
require that $M_{top}>46$\ \GeVcc\ at 95\%\ confidence level (CL). 
These limits are insensitive to the decay modes of the top quark and
the coupling of the top quark to the electroweak bosons.

Another relatively model-independent limit is set by measurements of
the width of the $W^+$\ boson.  
Direct and indirect measurements \cite{ref: Gamma W measurements}\
of $\Gamma_W$\
indicate that the top quark is massive enough that the decay channel
$W^+\rightarrow t\bar{b}$\ does not contribute to $\Gamma_W$.  The
limit set is $M_{top}>62$\ \GeVcc\ at 95\%\ CL.  

Direct searches for the top quark at hadron colliders have focused on
two specific models for top quark decay: i) the minimal
supersymmetric model (MSSM)\cite{ref: mssm}\ where the decay mode $t\rightarrow H^+ b$\
is also allowed ($H^+$\ is the charged Higgs boson), and 
ii) the standard model where the
top quark decays directly to $t\rightarrow W^+b$.  The most stringent
limit\cite{ref: charged Higgs}\ assuming the MSSM requires that $M_{top}>96$\
\GeVcc\ at 95\%\ CL for the case
where $t\rightarrow H^+b$\ always and 
$BR(H^+\rightarrow \tau^+\nu_\tau)=1.0$.
This limit, however, depends on the overall width of the
decay  $t\rightarrow H^+ b$, the Higgs boson branching fractions
($H^+$\ is expected to preferentially decay to $c\bar{s}$\ and
$\tau\nu_\tau$\ final  states) and
the $H^+$\ detection efficiency.
The \DZero\ collaboration has published the most sensitive standard
model search  using a 15~\invpb\ dataset, and has excluded a  top quark
with mass less than 131
\GeVcc\ at 95\%\ CL \cite{ref: Dzero limit}.

On the other hand, the CDF collaboration published a study
of $\sim20$\ \invpb\ of data in April 1994 that claimed
evidence for top quark production \cite{ref: CDF PRD/PRL}.
A total of 12 events were observed in several decay modes above
a predicted background of approximately 6 events.  The probability
that the observed event rate was consistent with a background
fluctuation was estimated to be 0.25\%.  In addition, evidence was
presented that the events in the sample were consistent with
arising from the production and decay of a $t\bar{t}$\ system
and inconsistent with the properties expected of the dominant 
backgrounds. 
Although compelling, this observation was statistically limited and
the possibility that it arose from a background fluctuation could
not be ruled out. 

In this report, I will focus on the latest results to come
from the \DZero\ and CDF top quark searches using data collected
between 1992 and 1995.  Both collaborations
have acquired over three times more data, and have
now reported conclusive evidence for top quark production 
\cite{ref: CDF/D0 Run Ib Results}.  I will
describe the analyses performed by both collaborations
and compare the two results.  

I believe an extremely persuasive
case has been made that the top quark has been found.
  
\section{Production and Decay of Heavy Top}
The production of heavy quarks in 1.8 TeV proton-antiproton
($p\bar{p}$) collisions is
predicted to take place through the two leading-order
quantum-chromodynamic (QCD) diagrams
\begin{eqnarray}
q\bar{q} &\rightarrow& Q\bar{Q} \label{eq: qqtoQQ} \\
gg &\rightarrow & Q\bar{Q}, \label{eq: ggtoQQ}
\end{eqnarray}
with the relative rate of these two processes dictated largely by the
mass of the heavy quark ($Q$), the parton
distribution functions of the proton and phase space.  Top quark
pair-production is expected to dominate the production rate. The
production of
single top quarks through the creation of a virtual $W^+$\ is 
smaller\cite{ref: single top}\ (of order 10\%\ of the \ttbar\ rate) 
and expected to occur in a relatively small part of
phase space.  All heavy top quark searches have therefore ignored single
top production. 

\begin{figure}
\vspace{4.5in}
\vskip 1in
\hskip 0.25in
\includegraphics{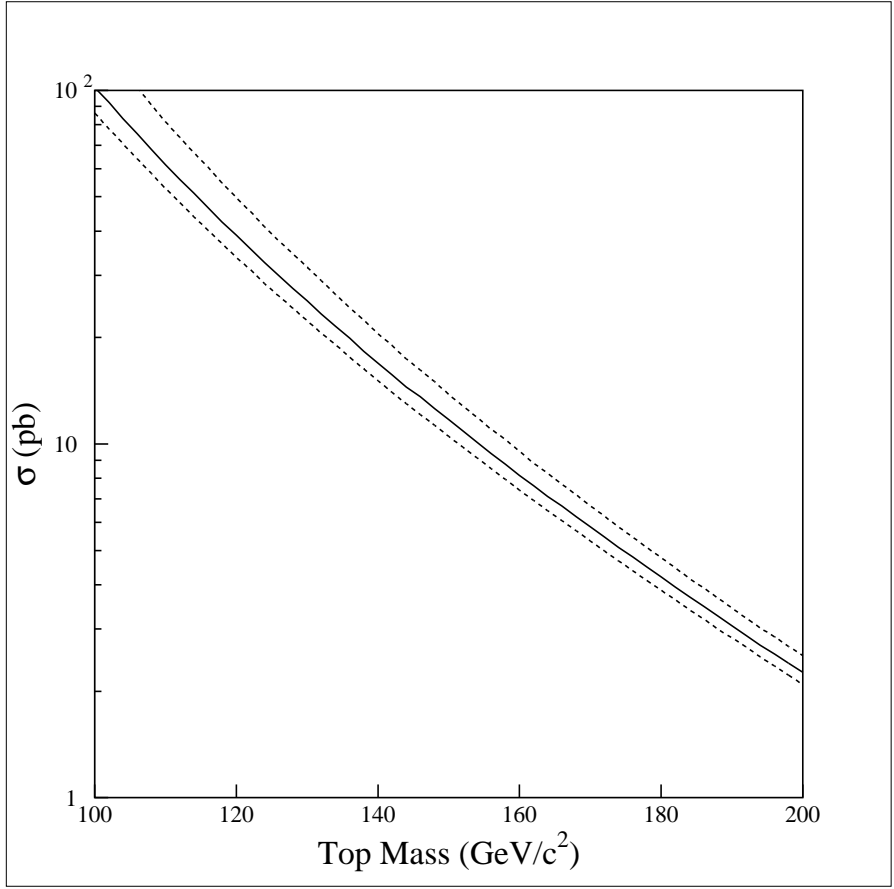}
\vskip -1in
\caption{The total cross section for top quark production in 1.8 TeV
$p\bar{p}$\ collisions as estimated by E.~Laenen \etal.
The upper and lower curves are a measure of the theoretical uncertainties
in the calculation.
}
\label{fig: Q cross section}
\end{figure}

The next-to-leading order corrections \cite{ref: Q sigmas}\ to 
processes (\ref{eq: qqtoQQ}) and (\ref{eq: ggtoQQ}) are relatively small
for heavy quark masses greater than $\sim 50$~\GeVcc.
More recently, these estimates have been
revised taking into account the effects of internal soft-gluon 
emission \cite{ref: soft-gluon corrections,ref: more on soft-gluons}.  
These cross sections
are shown in Fig.~\ref{fig: Q cross section}\ plotted as a function
of the heavy quark mass.  The uncertainty in these estimates reflects
the theoretical uncertainty in this calculation, which is
believed to be the choice of renormalisation scale.  
For top quark masses above 100 \GeVcc, the primary contribution to the
cross section comes from quark annihilation. 
This
reduces the uncertainties arising from our lack of knowledge of
the parton distribution functions of the proton, as these have 
been relatively accurately measured at large Feynman $x$, the kinematic region
that would dominate very heavy quark production.

Top quark pair production will generate a top quark and
anti-top quark that are recoiling against each other in the lab.  The
production diagrams favour configurations where the top quarks
are produced isotropically in the lab frame.  The relative motion of
the \ttbar\ system is expected to be small in comparison to the
transverse momentum\footnote{
I will employ a coordinate system where the proton beam direction defines
the $\hat{z}$\ axis, and transverse variables such as transverse
momentum (\Pt) and transverse energy (\Et) are defined relative to this
axis.  The angle $\phi$\ represents the azimuthal angle about the
beam axis and the angle $\theta$\ represents the polar angle relative to 
the beam axis.  Pseudorapidity $\eta\equiv -\ln\tan(\theta/2)$\ will often
be employed instead of $\theta$.}\
 (\Pt) distribution of the top quark 
itself \cite{ref: ttbar kinematics}.  The expected \Pt\
distribution for a heavy top quark has a peak around half
the top quark mass with a relatively long tail. The pseudorapidity
distribution for top quarks is peaked at 0 and falls off rapidly
so that most of the top quarks are produced in the ``central'' region
with pseudorapidity $|\eta|<2$.
The combination of a relatively
energetic heavy quark produced centrally is ideal from an
experimental point of view.  The top quark decay products are 
rather stiff and central, aiding their detection.

The standard model predicts that
the top quark will decay almost always via $t\rightarrow W^+ b$.
The $W^+$\ decays approximately 2/3 of the time into
$q\bar{q}^\prime$\ pairs  ($u\bar{d}$\ or $c\bar{s}$) and 1/3 of the
time into one of the three lepton generations.  This results in a
decay topology consisting of 6 energetic partons that could either be
charged leptons, neutral leptons or quark jets.  

The decay channels involving $\tau^+$\ leptons
are problematic given the difficulty of cleanly
identifying these weakly decaying leptons in a hadron collider environment.  They
have therefore not been explicitly included in the searches I describe
below.  The final states involving 6 quark jets suffer an enormous
background from QCD multijet production, with estimates of intrinsic
signal-to-noise of $<10^{-4}$.  
Because of these large backgrounds, this channel has not been the
focus of most of the effort, and I will ignore it here also.
However, recent work has demonstrated that a significant \ttbar\ signal
can be observed in these modes.\cite{ref: ttbar in jets}

With these considerations, there are five final states that are
experimentally accessible:  
\begin{eqnarray}
\mttbar &\rightarrow& e^+\nu_e b\     e^-\bar{\nu_e} \bar{b} 
\quad (1/81)
\nonumber \\
\mttbar &\rightarrow& \mu^+\nu_\mu b\ \mu^-\bar{\nu_\mu} \bar{b} 
\quad (1/81)
\nonumber \\
\mttbar &\rightarrow& e^+\nu_e b\   \mu^-\bar{\nu_\mu} \bar{b}
\quad (2/81)
\label{eq: fifth channel} \\
\mttbar &\rightarrow& e^+\nu_e b\    q\bar{q}^\prime \bar{b} 
\quad (12/81)
\nonumber \\
\mttbar &\rightarrow& \mu^+\nu_\mu b\ q\bar{q}^\prime \bar{b}
\quad (12/81),
\nonumber
\end{eqnarray}
where I have also listed the expected standard model branching fractions
for each channel. In all cases where I refer to a specific charge state,
the charge conjugate mode is implied.  
The first three dilepton channels turn out to be
the cleanest final states, as the requirement of two energetic
charged leptons and neutrinos virtually eliminates all 
backgrounds.  They suffer from rather small branching fractions
and are therefore the most statistically limited.  The last two
lepton+jets final states together correspond to approximately 30\%\
of the \ttbar\ branching fraction.  However, these channels
face the largest potential backgrounds.

\section{Backgrounds to Top Quark Searches}
Top quark production is an extremely rare process in \pbarp\
collisions; its cross section of less than 100 pb can be compared
with the total \pbarp\ cross section of over 50 mb (almost nine
orders of magnitude difference).  Since the total cross section is
dominated by ``soft'' QCD interactions, the top quark cross section
can be more fairly compared with the cross section for other high
$Q^2$\ production processes, such as inclusive
$W^+$\ production (20 nb), $Z^\circ$\ production (2 nb) and 
$W^+W^-$ and $W^+Z^\circ$
production (10 and 5 pb, respectively).  These processes
are the sources of the most severe background to \ttbar\ production.

It is necessary to control these backgrounds so that one can be
sensitive to a top quark signal. All the channels listed in
Eqs.~(\ref{eq: fifth channel})\ involve an
energetic charged electron or muon, and one or more energetic
neutrinos.  The requirement of these two signatures in the final state
using the
\DZero\ and CDF lepton identification systems are sufficient to
adequately control the backgrounds associated with jets that might
satisfy the lepton ID criteria.  The remaining backgrounds are
dominated by physics processes that generate real leptons in the final
state.  

In the case of the dielectron and dimuon modes, the single largest
background comes from Drell-Yan production (including
$Z^\circ\rightarrow e^+e^-$\ and $Z^\circ\rightarrow\mu^+\mu^-$).  This is
controlled by requiring a neutrino signature as well as
additional jet activity.  The single largest physics
background in the $e^+\mu^-$\ final state comes from
$Z^\circ\rightarrow\tau^+\tau^-$\ decay, which can be similarily reduced by
the requirement of a neutrino signature and additional jets.

The single largest physics background to lepton+jets final states
come from inclusive $W^+$\ production where additional jets are
produced via initial and final state radiation 
\cite{ref: VECBOS calculations}.  The intrinsic rate for this
background depends strongly on the multiplicity 
requirements placed on the jet candidates, as shown in 
Table~\ref{tab: W+jet rates}\ where the observed $W$+jet production cross
section is presented as a function of jet
multiplicity and compared 
with a QCD Monte Carlo prediction \cite{ref: Observed W+jet sigma}.
\begin{table}
\begin{center}
\begin{tabular}{ccc}
\hline
Jet Multiplicity	& $\sigma B$\ (pb)  & $\sigma_{T}B$\ (pb) \\
\hline
0		& $1740 \pm 31\pm288$		&  $1753 \pm 26\pm123$ \\
1		&  $336 \pm 14 \pm 63$    	&   $287\pm4\pm21$ \\
2		&  $76 \pm 12 \pm 18$		&    $59 \pm 2 \pm 5$ \\
3		&  $ 14\pm 3 \pm 3$		& $11.0\pm0.3\pm1.0$ \\
4		&  $4.0\pm1.6\pm1.2$		& $2.0\pm0.1\pm0.3$ \\
\hline
\end{tabular}
\end{center}
\caption{The $W$+jet production cross section times the branching ratio
for $W^+\rightarrow l^+\nu_l$\ as a function of jet multiplicity.
The second column presents the observed cross sections for jets with
corrected transverse energy $>15$\ GeV and $|\eta|<2.4$.  
The third column shows the
predicted QCD cross section based on a VECBOS Monte Carlo calculation.
} 
\label{tab: W+jet rates}
\end{table}
One can see from these rates that this background can 
overwhelm a \ttbar\ signal. More stringent kinematic cuts can be
applied to reject the $W$+jet events, taking advantage of the fact
that the \ttbar\ final states, on average, generate higher \Et\
$W^+$\ bosons and additional jets.  Alternatively, since the \ttbar\
final state has two $b$\ quark jets in it, the requirement that one or
more jets are consistent with arising from the fragmentation and decay 
of a $b$ quark will preferentially reduce the
$W$+jets background.  Both of these techniques have been employed.

\section{The Tevatron Collider}
The Tevatron Collider is a 6 km circumference proton-antiproton
storage ring that creates \pbarp\ collisions at a centre-of-mass
energy of 1.8 TeV.  In its current configuration, the collider
operates with six bunches of protons and six bunches of
counter-rotating antiprotons that are brought into collision at two
intersection points in the ring named B0 and D0.  The B0
and D0 interaction regions house the CDF and \DZero\ detectors,
respectively.

The Tevatron embarked on a multi-year collider run starting in
December 1992.  The first stage of the run, known as Run IA,
continued till August 1993, at which time approximately 30
\invpb\ had been delivered to each interaction region.  The second
stage, Run IB, commenced in August 1994 and
by February 1995 the collider had delivered an
additional 80 \invpb\ to each interaction region.  The maximum
luminosity of the Collider during this period was $1.7\times10^{31}$\ ${\rm
cm^{-2}s^{-1}}$, and has been steadily rising.

Run IB run ended in February 1996, with a
total of $\sim150$\ \invpb\ delivered to each interaction region.

\section{The \DZero\ and CDF Experiments}
The \DZero\ and CDF detectors have been designed to trigger and
record the high \Pt\ collisions that result when two partons in the
\pbarp\ system undergo a hard scatter.  Both instruments 
detect electrons, muons, neutrinos and quark and gluon jets using
a set of complementary subdetectors.  However, they accomplish this
common goal in rather different ways.

\subsection{The \DZero\ Detector}
\begin{figure}
\vspace{4.5in}
\vskip 1.2in
\hskip 0.5in \includegraphics{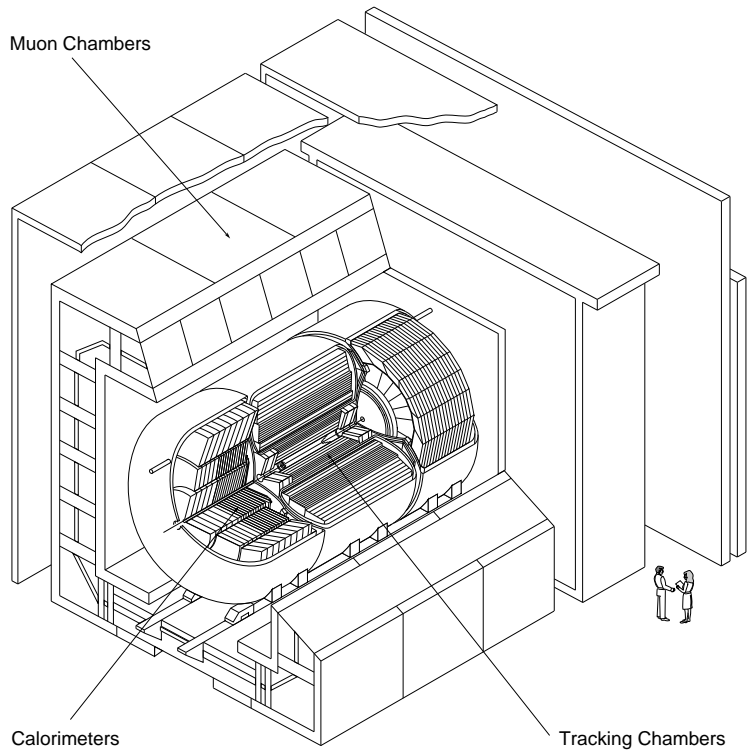}
\vskip -1.2in
\caption{A cut-away view of the \DZero\ detector.
The inner tracking detectors are surrounded by the calorimeter cryostats,
and both are situated inside the toroidal magnet.  Planes of chambers 
outside the magnet provide muon identification and momentum
measurement.
}
\label{fig: Dzero schematic}
\end{figure}

The \DZero\ detector was designed with the philosophy that a
uniform, hermetic, highly-segmented calorimeter should form the
core of the detector \cite{ref: D0 detector}.  
A cut-away view of the detector is shown in
Fig.~\ref{fig: Dzero schematic}. The
\DZero\ calorimeter employs a uranium absorber up to nine interaction
lengths thick and a liquid argon readout system.  This provides
excellent hermeticity and uniformity, except perhaps in the
transition region between the barrel and endcap cryostats.  
The overall
resolution of the \DZero\ calorimeter is 
\begin{eqnarray}
{{\sigma_E}\over{E}} &=& {{0.15}\over{\sqrt{E}}} \oplus 0.004 \
\mbox{for\ electromagnetic\ showers} \\
{{\sigma_E}\over{E}} &=& {{0.80}\over{\sqrt{E}}} \
\mbox{for\ hadrons},
\end{eqnarray}
where $E$\ is measured in GeV.

A muon
system consisting of charged particle detectors and 1.9 Tesla toroidal
magnets located outside the calorimeter provides good muon
identification.  
This system identifies muon candidates in the region $|\eta|<3.3$\ 
using sets of
muon tracking chambers consisting of proportional drift tubes
located interior and exterior to the large toroidal magnetic field.  The
deflection of the muon candidates in the magnetic field provides a 
momentum measurement with an accuracy of
\begin{eqnarray}
\sigma\left( {{1}\over{p}}\right) = {{0.18\left(p-2\right)}\over{p^2}} \oplus 0.008,
\end{eqnarray}
where $p$\ is the muon momentum measured in \GeVc.

Vertex, central and forward drift chambers provide
charged particle detection in the interval $|\eta|<3.2$.
The tracking system does
not incorporate a magnetic field, as the presence of a magnetic coil
would degrade calorimeter performance.  

\subsection{The CDF Detector}
\begin{figure}
\vspace{3.5in}
\vskip 6.0in
\hskip-0.9in
\includegraphics{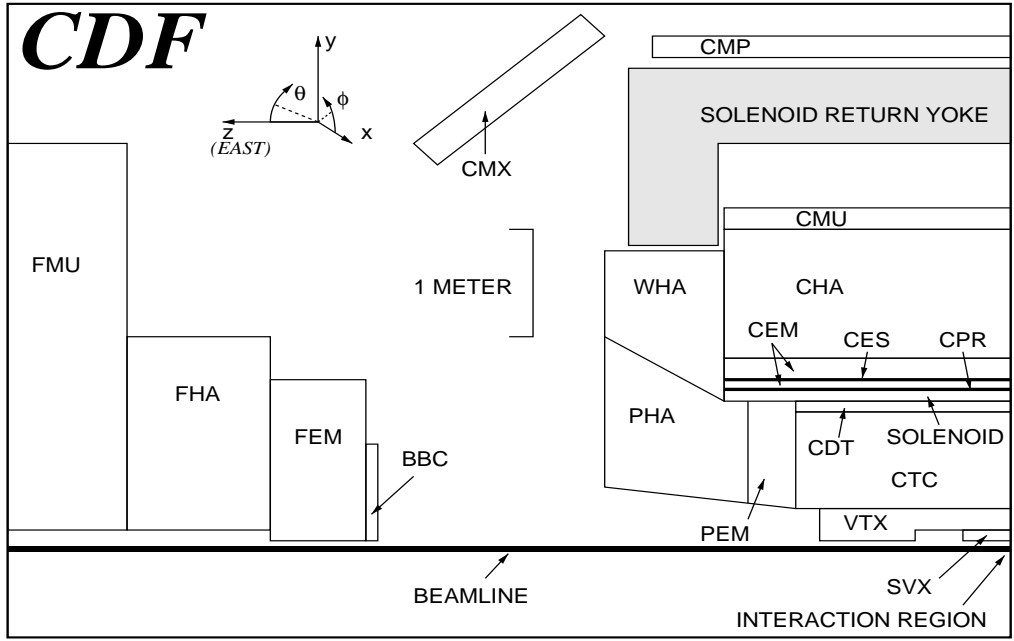}
\vskip-6.0in
\caption{A schematic view of one quarter of the CDF detector. 
The interaction point is at the lower right corner of the figure.
}
\label{fig: CDF schematic}
\end{figure}

The CDF detector \cite{ref: CDF detector}\ consists of a high-precision tracking
system in a 1.4 T solenoid magnetic field, surrounded by a hermetic
highly-segmented calorimeter, as shown in 
Fig.~\ref{fig: CDF schematic}.  The tracking system consists of three
independent devices arranged coaxial to the beam line.  A 4-layer
silicon-strip detector (SVX) with inner and outer radii of 3.0 and 7.9 cm
provides of order 40
$\mu$\ precision on the impact parameter of 
individual charged track trajectories extrapolated
to the beam line.  A set of time projection chambers (VTX) instrument the
tracking region between 12 and 22 cm in radius, providing
high-precision tracking in the $r$-$z$\ plane.  An 84-layer drift
chamber (CTC) detects charged particles in the region between 30 and 132 cm
from the beamline.  Together, these detectors measure particle
transverse momentum to a precision $\sigma_{p_T}$\ given by
\begin{eqnarray}
{{\sigma_{p_T}}\over{p_T}} &=& 0.0009p_T \oplus 0.0066,
\end{eqnarray}
for particles with $p_T \gessim 0.35$\ \GeVc.  

The central calorimeter (CEM and CHA)
instruments the region $|\eta|<1.1$, and is comprised of projective
towers of size
$\Delta\eta\times\Delta\phi=0.1\times0.26$\ radians. Each tower
is made of a sandwich of Pb or Fe plates interleaved with
scintillator.  A Pb sandwich 25 radiation lengths thick is used
to measure electromagnetic shower energies.  
An iron-scintillator sandwich approximately 5 interaction 
lengths
thick is used to detect hadronic showers.  Plug and Forward
calorimeters (PEM, PHA, FEM and FHA)
instrument the region $1.1 < |\eta| < 4.2$, and consist
of similar absorber material.  The showers in this region are detected with
proportional wire chambers as they provide for a more
radiation-resistant detector system.   The presence of a solenoid magnet
and a significant amount of material in front of the calorimeter leads
to some compromise in calorimeter performance.
The overall resolution of the CDF calorimeter is
\begin{eqnarray}
{{\sigma_E}\over{E}} &=& {{0.137}\over{\sqrt{E}}} \oplus 0.02 \quad
\mbox{(for\ electromagnetic\ showers)}\\
{{\sigma_E}\over{E}} &=& {{0.50}\over{\sqrt{E}}} \oplus 0.03 \quad
\mbox{(for\ hadrons)}.
\end{eqnarray}

Planar drift chambers (CMU, CMP and CMX) located outside the calorimeter volume
detect muons penetrating the calorimeter absorber, but precise muon
momentum and direction  come from the associated charged track
detected in the inner tracking system.  The central muon system is
able to detect muons within the pseudorapidity interval
$|\eta|<1.0$.  A forward muon system (FMU) consisting of large toriodal
magnets surrounded by drift chambers and scintillator counters 
detect muons in the rapidity region $2.2\le|\eta|\le 3.5$.  

\subsection{Triggering and Data Acquisition}
 
Pair production of standard model top quarks and their subsequent
decay into either the dilepton or lepton+jets mode yields a signature
that is relatively straightforward to trigger on.  Both detectors
employ multi-level trigger systems where at each level more
information is brought together to form a decision.  The trigger
requirement of at least one energetic electron or muon is the primary
tool used in identifying online a sample of top quark
candidate events that are subsequently studied offline.

The requirement of at least one high
\Pt\ electron or muon in both CDF and
\DZero\ is imposed efficiently in the trigger.  The
production of leptons above a transverse energy of 15 GeV is dominated
in both experiments by $b$\ and $c$~quark production, and by
inclusive
$W^+$\ boson production.   For example, in CDF, the inclusive
electron trigger is implemented with the following requirements:
\begin{enumerate}
\item The level 1
trigger demands that at least one calorimeter trigger cell with
$\Delta\phi\times\Delta\eta = 0.26\times0.2$\ has $>6$~GeV
of electromagnetic energy. 
\item The level 2 trigger demands that there be a 
charged track candidate pointing at an electromagnetic energy
cluster, and requires that the cluster properties be consistent with
those of an electromagnetic shower.
\item The level 3 trigger requires the presence of an electromagnetic
cluster associated with a charged track reconstructed using the standard offline
algorithms.  Further quality cuts on the
properties of the electromagnetic shower are also made.
\end{enumerate}
These reduce the overall cross section of candidate events to
approximately 50 nb, of which approximately 30\%\ is comprised
of real electrons.  For comparison, the rate of $W^+\rightarrow
e^+\nu_e$\ in this sample is of order 1 nb.  The efficiency of this 
trigger for isolated electrons with $20 < E_T < 150$~GeV is $92.8\pm0.2$\%.

As another example, the \DZero\ detector triggers on a sample of
inclusive muon candidates by using a two level decision process:
\begin{enumerate}
\item The level 1 trigger demands the presence of a charged track stub
in the muon toroidal spectrometer with a $p_T>3$~\GeVc.
\item The level 2 trigger demands a high quality muon candidate consisting
of a muon candidate in the muon system matched to a charged track 
observed in the central tracking system.
The central track candidate must be reconstructed in all 3 dimensions,
must be consistent with coming from the event interaction and must have
\Pt\ greater than 5 or 8 \GeVc, depending on the specific muon trigger.
\end{enumerate}
The efficiency of this trigger is estimated to be
$67\pm3$\%.

Both experiments employ inclusive electron and muon triggers, as
well as triggers that identify smaller samples of events useful to
the top search.  Since the backgrounds to the dilepton sample are
relatively small, it is convenient to identify the candidate events
immediately in the trigger so that they can be analysed as soon as
possible.  A high-\Pt\ dilepton trigger requiring
at least two electron or muon candidates is therefore employed to
flag these candidates immediately.  The cross section for this 
trigger is only a few nb.

At a luminosity of $2\times 10^{31}$~\invcms, a trigger
cross section of 300~nb corresponds to an event rate of 6~Hz, which can
be comfortably recorded and analyzed.  Note, however, that even with
a cross section of 10~nb, the total data sample for an integrated
luminosity of 50~\invpb\ will consist of 500~000 events, with each
event comprised of order 200 kbytes of information.  

\subsection{The Run IA and IB Datasets}
The Tevatron Collider started up after a three year shut-down in fall
1992, and continued running through the summer of 1993.  
As this was the \DZero\
detector's first collider run, it was remarkable that the
collaboration was able to successfully use 40-50\%\ of the collisions
for their physics studies.  The CDF collaboration gathered $19.6\pm0.7$\
\invpb\ of data during this period.

From the start of Run IB in 1994 to February 1995, the Tevatron Collider
had delivered over 100 \invpb\ of collisions to each
detector.  The 
\DZero\ and CDF collaborations had recorded and analysed $\sim45$\
\invpb\ of this data by this date, giving the the two collaborations
total Run I datasets of 50 and 67~\invpb, respectively.  

In between Run IA and IB, both collaborations made incremental
improvements to their detectors.  The \DZero\ detector's muon trigger
was improved and various detector subsystems were modified with the
goal of improving overall robustness and efficiency.  The CDF
collaboration replaced the original 4-layer SVX detector with a
mechanically identical device that used newer,
radiation-hard silicon strip wafers, and employed an
AC-coupled readout design.  The new detector, known as the SVX', has
much better signal-to-noise and is fundamentally better understood. 

\subsection{Event Reconstruction}
\label{sec: kinematic variables} 
\begin{figure}
\vspace*{4.25in}
\vbox{
\vskip+0.0in
\includegraphics{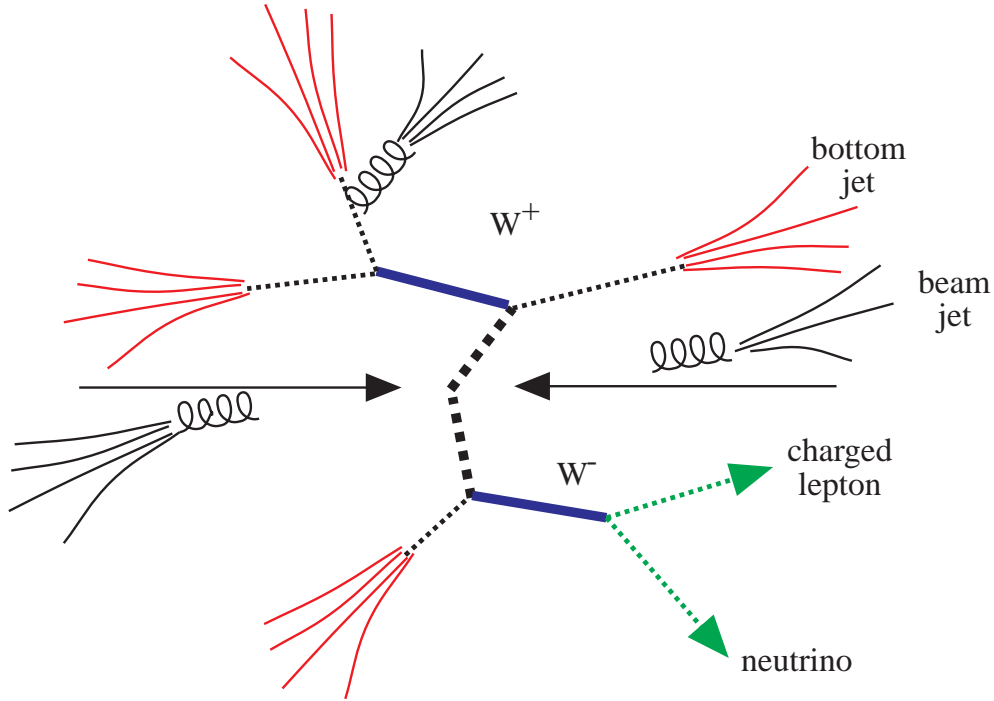}
\hskip-0.0in
\vskip-0.0in
}
\caption{A schematic of a \ttbar\ event produced at the Tevatron and
decaying into a lepton+jets final state.  In addition to the partons
resulting from the decay, there are additional jets produced by initial
and final state radiation.
}
\label{fig: ttbar schematic}
\end{figure}

A schematic of a \ttbar\ event being produced in a
\pbarp\ collision and decaying into the final state partons is
shown in Fig.~\ref{fig: ttbar schematic}\
Given the large number of partons that arise from the decay of the
\ttbar\ system, each detector is required to reconstruct with good
efficiency high energy electrons, muons and the jets resulting from
the fragmentation of high energy quarks, and to tag the presence of one or
more neutrinos by the imbalance of total transverse energy in the
collision.

High energy electrons and muons are identified in both
detectors by the charged track left in the central tracking systems,
and by the behaviour of the leptons in the calorimeters and muon
identification systems outside the calorimeters.  Electrons will
generate an electromagnetic shower in the calorimeter, with a lateral
and longitudinal shower profile quite distinct from the shower
intitiated by a charged hadron.  Muons are readily identified as they
generally pass unimpeded through the calorimeter and are detected
outside the calorimeters as charged particles that point back to the
particle trajectory in the central tracker.  The CDF electron
and muon reconstruction algorithms have efficiencies of $84\pm2$\%\
and $90.6\pm1.4$\%\ for leptons from $W^+$\ boson decays.  The \DZero\
electron reconstruction has an efficiency of $72\pm3$\%. 
These efficiencies are quoted for
electron and muon candidates that have already passed the trigger
requirements discussed earlier.

Neutrinos can only be detected by requiring that they have sufficient
transverse energy that the total measured energy flow sum to a
value inconsistent with zero.  In practical terms, this energy flow
vector is known as missing transverse energy (\MEt).  Note that we
cannot use the imbalance in energy flow along the beamline in this
case as one can expect a significant imbalance due to the differing
momentum of the partons in the proton and antiproton that collide to
produce the \ttbar\ system.  The resolution in \MEt\ is driven by both
the uniformity of the calorimeter and its inherent energy resolution. 
\DZero\ has a missing transverse energy resolution in each transverse
coordinate of
\begin{eqnarray}
\sigma_{x} = 1.08 + 0.019\left(\sum E_T\right)\ {\rm GeV},
\end{eqnarray}
where the summation gives the total scalar transverse energy observed
in the calorimeter.  CDF's transverse energy resolution is
approximately 15-20\%\ worse, which has a modest impact on its
neutrino detection ability.

Jets are constructed in both detectors as clusters of transverse
energy within a fixed cone defined in $\eta$-$\phi$\ 
space \cite{ref: jet cluster algorithm}.
The size
of this cone is determined by the competing requirements of making it
large enough to capture most of the energy associated with the
fragmentation of a quark or gluon, and yet small enough  that it
doesn't include energy associated with nearby high energy partons or
from the ``underlying'' event.  The latter effect in itself
contributes on average approximately 2 GeV per unit in $\eta$-$\phi$\ space, and
the fluctuations in this degrades the jet energy
resolution (the size of this effect depends on the rate of
multiple interactions).  
Monte Carlo (MC) calculations using a variety of models
for quark fragmentation and underlying event assumptions, as
well as studies of the underlying events have indicated that a jet
cluster cone size substantially smaller than the traditional 
$\eta$-$\phi$\ radii of 0.7 or 1.0 employed in QCD studies is required.
The CDF analysis employs a cone radius of 0.4 in its top quark
search, whereas the \DZero\ collaboration has chosen to work with a
cone radius of 0.5.  

\begin{figure}
\vspace{3in}
\vskip 6.5in
\hskip -1.5in
\includegraphics{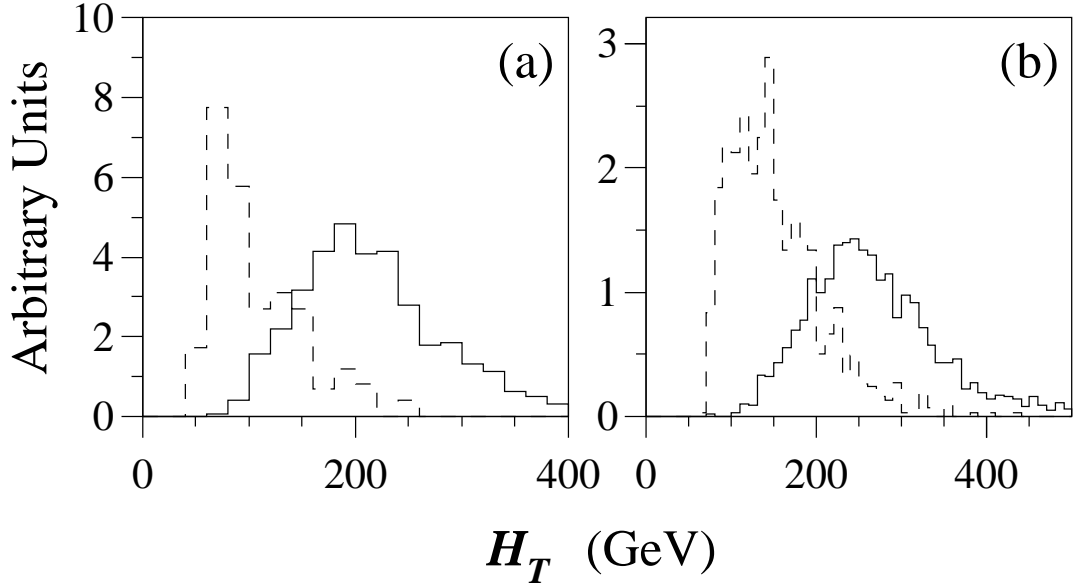}
\vskip -6.5in
\caption{The $H_T$\ distributions for $e^+\mu^- +$jet events (a) and
lepton+jet events (b).  The solid histograms are the distributions expected
from \ttbar\ events for a top quark mass  of 200 \GeVcc.
The dashed histograms are the expected distributions for the dominant
backgrounds to \ttbar\ production in both channels.
}
\label{fig: HT distributions}
\end{figure}

The requirement
that most if not all daughters are reconstructed is not sufficient
to reject all backgrounds to \ttbar\ production.  There are
other kinematical variables that
discriminate between \ttbar\ and background events, most of them
taking advantage of the fact that heavy top quark production will
generate final state daughters that are on average quite energetic.
This motivates the use of a variable called
$H_T$\ defined as
\begin{eqnarray}
H_T = \sum_{i=1}^{N_p} E_T^i,
\end{eqnarray}
where the sum is over all the jets and the leading electron cluster 
(in those channels where at least one electron is required).
This variable is used by the \DZero\ collaboration in both their
dilepton and lepton+jets analysis, and its effectiveness in improving
the signal-to-noise in the dilepton and lepton+jets channels is
illustrated in Fig.~\ref{fig: HT distributions}.  The CDF collaboration has
recently reported the results of a top analysis using a similar
variable \cite{ref: CDF H Analysis}.

An additional kinematic variable known as aplanarity
\cite{ref: aplanarity definition}\ 
({$\aplanarity$}) has been employed by the \DZero\ collaboration. 
This, as its name suggests,
is a measure of how spherical a candidate event is: \ttbar\ events
are expected to have larger values of $\aplanarity$\ than the corresponding
physical backgrounds.  

The final tool used in the reconstruction of \ttbar\ events is the
identification or ``tagging'' of jets that arise from the $b$\ quarks.
There are two techniques employed by the collaborations.  The first
takes advantage of the fact that bottom hadrons decay
semileptonically into electrons or muons about 20\%\ of the time.
\DZero\ and CDF therefore search the interior of each jet cone for a
muon candidate. CDF also searches
for low-energy electron candidates that can be associated with the jet
cluster.  Because there are two $b$\ quarks in each \ttbar\ decay,
the efficiency of this soft lepton (SLT) tagging scheme ranges from 10-15\%.
The second technique is used exclusively by CDF and takes advantage of
the long-lived nature of bottom hadrons and the SVX (or SVX')
detector.  A seach is performed for several charged tracks 
detected in the SVX that form a secondary
vertex a significant distance from the primary interaction. 
The efficiency of this tagging scheme
depends crucially on the performance of the SVX/SVX'.  It is
estimated that over 40\%\ of all \ttbar\ decays will have the
presence of at least one SVX tag.

\section{The Dilepton Top Quark Search} 
\subsection{Dilepton Data Selection}
The dilepton decay modes are the cleanest
channel in which one would expect to observe a heavy top quark.  
They suffer from the relatively small total branching fraction of
\ttbar\ into these modes (a total of 4\%), and from the presence of
two neutrinos in the final state that are not individually observable.

The dilepton searches break down into three separate channels,
the $e^+e^-$, $\mu^+\mu^-$\ and $e^+ \mu^-$\ final states.
The CDF analysis requires two isolated lepton candidates, each with
$P_T>20$\
\GeVc\ and with $|\eta|<1.0$.  The candidates must satisfy
standard lepton quality requirements that ensure high efficiency and
high rejection from energetic, isolated charged hadrons.
There are 2079 $e^+e^-$\ candidates, 2148 $\mu^+\mu^-$\ candidates and 
25 $e^+\mu^-$\
candidates after these kinematical cuts.  
The large $e^+e^-$\ and $\mu^+\mu^-$\ candidate samples are the result of
$Z^\circ$\ and Drell-Yan production, as can be seen by examining
the invariant mass ($M_{ll}$) distribution of the dilepton system.
This background is removed by rejecting those events with 
\begin{eqnarray}
75 < M_{ll} < 105\ \mGeVcc.
\end{eqnarray}
This leaves 215, 233 and 25 candidate events in the 
$e^+e^-$, $\mu^+\mu^-$\ and $e^+\mu^-$\ 
channels, respectively.

\begin{figure}
\vspace*{6.0in}
\vskip 1.5in
\includegraphics{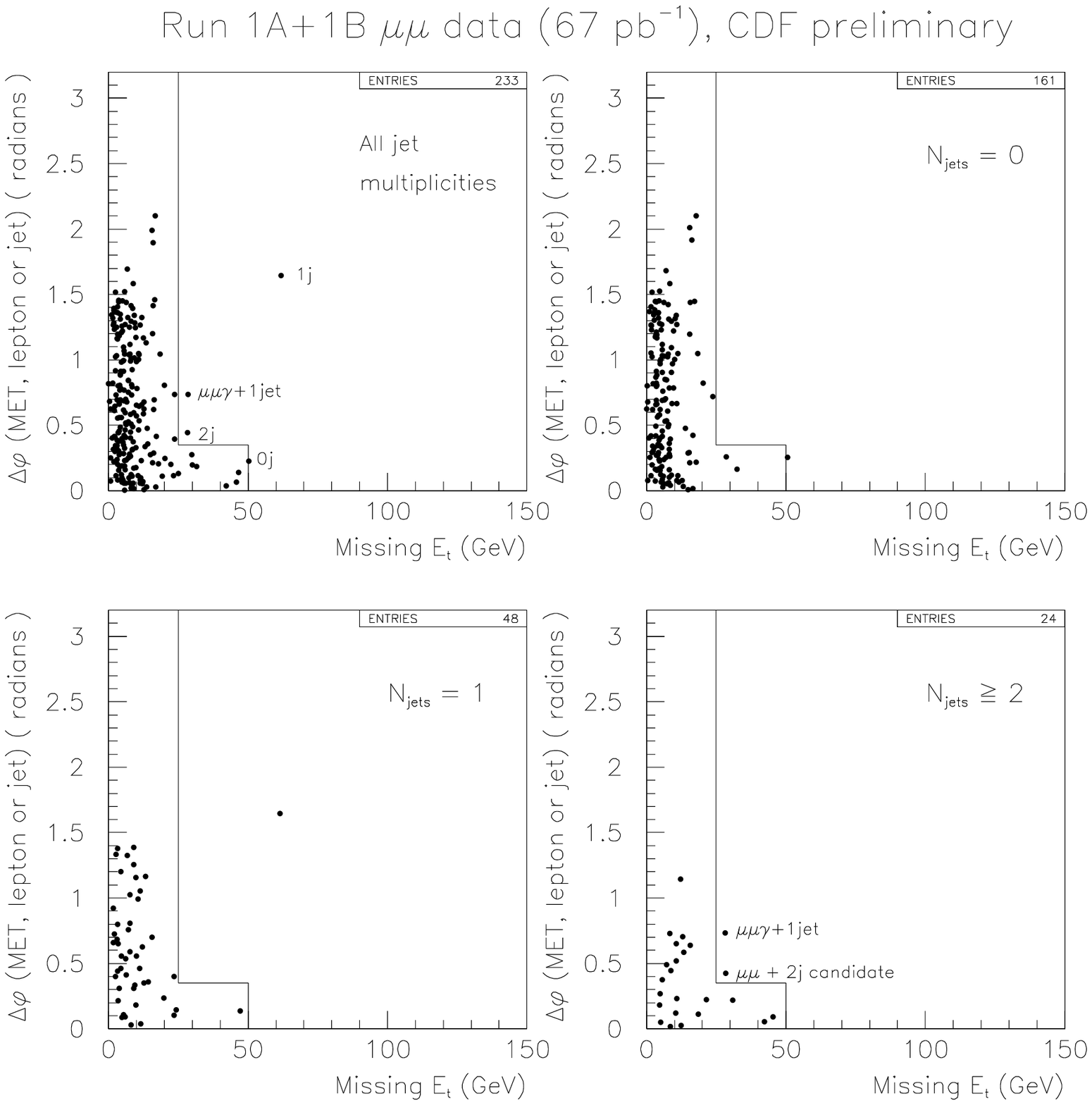}
\vskip -1.5in
\caption{The distribution of the azimuthal opening angle between
the missing \Et\ vector and the highest energy jet or lepton versus
the \MEt\ is shown for all CDF candidate events, and for events with
0, 1 and $\ge2$\ jets in the $\mu^+\mu^-$\ channel.
The boundary shows the
cuts placed to reject the remaining Drell-Yan background. }
\label{fig: dphi vs met mumu}
\end{figure}

\begin{figure}
\vspace*{6.0in}
\vskip 1.5in
\includegraphics{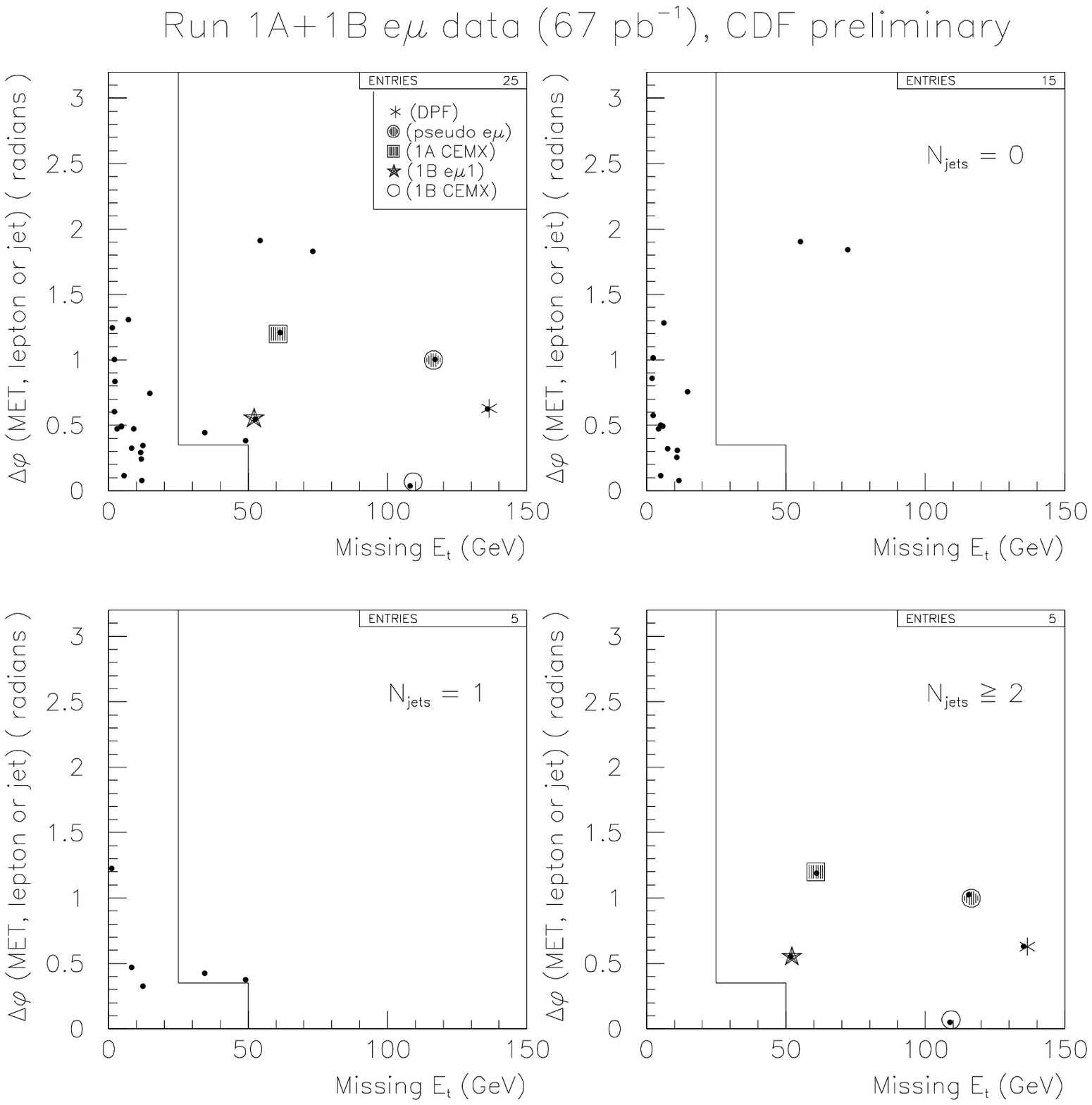}
\vskip -1.5in
\caption{The distribution of the azimuthal opening angle between
the missing \Et\ vector and the highest energy jet or lepton versus
the events \MEt\ is shown for all CDF candidate events, and for events with
0, 1 and $\ge2$\ jets in the $e^+\mu^-$\ channel.
The boundary shows the
cuts placed to reject the $Z\rightarrow \tau^+\tau^-$\ background. }
\label{fig: dphi vs met emu}
\end{figure}

In addition,
the events are required to have $\mMEt>25$\ GeV and at
least two jet clusters with $E_T > 10$\ GeV and $|\eta|<2.0$, since
\ttbar\ events are expected to have two energetic neutrinos and 
a $b$~quark and anti-quark in the final state.
This still leaves a background in the $e^+e^-$\ and $\mu^+\mu^-$\ sample from
Drell-Yan production where the \MEt\ signal arises from an
accompanying jet that is mismeasured.  The distributions of the
azimuthal opening angle between the missing transverse energy vector and the
closest jet or charged lepton candidate in the event versus the
missing transverse energy for each jet multiplicity are shown in 
Figs.~\ref{fig: dphi vs met mumu}\ and
\ref{fig: dphi vs met emu}\
for the $\mu^+\mu^-$\ and $e^+\mu^-$\ channels, respectively.  
There is a clear cluster of events at small
\MEt-jet opening angles that extend to higher \MEt\ in the 
$\mu^+\mu^-$\ (and $e^+e^-$) samples that results from the remnant
Drell-Yan contamination in the samples.  The same enhancement is not present
in the $e^+\mu^-$\ sample, which has no Drell-Yan contamination.  
A stiffer \MEt\ cut requiring at least 50 GeV of missing transverse
energy is imposed on those events that have \MEt-jet opening angles
less than $20^\circ$.  The same region is occupied preferentially
by backgrounds from $Z\rightarrow \tau^+\tau^-$\ in the $e^+\mu^-$\ 
sample so it is also removed.

This leaves a total of 7 candidate CDF events, 5 in the $e^+\mu^-$\
channel and two in the $\mu^+\mu^-$\ channel.  No dielectron events
survive the selection.  
One of the $\mu^+\mu^-$\ events has an energetic photon candidate with a
$\mu^+ \mu^-\gamma$\ invariant mass consistent with that of a
$Z^\circ$\ boson. Although the expected background from radiative
$Z_\circ$\ decay is only 0.04 events,  the $\mu^+\mu^-\gamma$\
candidate is removed from the sample in order to be conservative.  

The \DZero\ analysis requires two high \Pt\ leptons; both leptons are
required to  have $P_T>20$\ \GeVc\ in the $e^+e^-$\ channel, 
$P_T>15$\ \GeVc\ in the $\mu^+\mu^-$\ channel, and $P_T>15(12)$\ \GeVc\
for the electron (muon) in the $e^+\mu^-$\ channel.  A \MEt\ cut
requiring at least 20 GeV and 25 GeV is placed on the $e^+\mu^-$\ and
$e^+e^-$\ channels, respectively (no \MEt\ requirement is placed on
$\mu^+\mu^-$\ candidate events).  The selection requires at least two
jets with corrected transverse energy $>15$\ with $|\eta|<2.5$.
Finally, $e^+e^-$\ and $e^+\mu^-$\ candidate events are required to have $H_T
>120$\ GeV and $\mu^+\mu^-$\ events are required to have $H_T>100$\ GeV.

This leaves a total of 3 dilepton candidate events in the \DZero\ 
dataset.  There are 
2 $e^+\mu^-$\ events, no $e^+e^-$\ events, and 1 $\mu^+\mu^-$\ event.  The 
integrated luminosities corresponding to these three channels is
$47.9\pm5.7$, $55.7\pm6.7$\ and $44.2\pm5.3$\ \invpb, respectively.
The number of observed events expected from \ttbar\ production
is shown in Table~\ref{tab: expected dilepton yield}.

\begin{table}
\begin{center}
\begin{tabular}{ccc}
\hline
Mass (\GeVcc)  &  \DZero & CDF \\
\hline
150      & 2.4     &  6.2  \\ 
160      & 2.0     &  4.4  \\ 
170      & 1.6     &  3.0  \\ 
180      & 1.2     &  2.4  \\
\hline
\end{tabular}
\end{center}
\caption{The expected number of dilepton events arising from \ttbar\
production for the \DZero\ and CDF selections as a function
of top quark mass.  The uncertainties on
these yields are of order 25-30\%.
The central value for the theoretical
prediction for the \ttbar\ cross section is assumed.}
\label{tab: expected dilepton yield}
\end{table}

\subsection{Dilepton Backgrounds}
The number of dilepton events observed by CDF and \DZero\ is
consistent with the rate expected from \ttbar\  production for a top
quark mass of order 140 to 150 \GeVcc.  It is necessary to accurately
estimate the number of events expected from standard model
background processes in order to interpret these event rates.

The most serious
potential background comes from $Z^\circ$\ boson production followed
by the decay $Z^\circ\rightarrow\tau^+\tau^-$. The $\tau^+$\ leptons then decay
leptonically leaving the dilepton signature and missing energy from
the four neutrinos.  The rate of this background surviving the 
selection criteria can be accurately estimated using the observed
$Z^\circ$\ boson kinematics in the dielectron and dimuon channels and
simulating the decay of the $\tau^+$\ leptons.  Other standard model
sources of dileptons are divector boson production, $b\bar{b}$\ and
$c\bar{c}$\ production and Drell-Yan production.  Most of 
these are either very small (e.g., the backgrounds from $W^+W^-$\ and
$W^+Z^\circ$\ production) or
can be estimated reliably from collider data (e.g. heavy quark
production).
Jets misidentified as leptons are a background source that also can be
accurately estimated using the data.  CDF uses the strong
correlation between fake lepton candidates and the larger energy flow
in proximity to the candidate.  \DZero\ employs similar techniques to
estimate this background.

The estimated background rates in the three channels are listed
in Table~\ref{tab: dilepton backgrounds}\ and total to $1.3\pm0.3$\ and
$0.65\pm0.15$\ for the CDF and \DZero\ analyses, respectively.  In both
cases, there is an excess of observed candidate events
above the expected backgrounds.

\begin{table}
\begin{center}
\begin{tabular}{lcc}
\hline
Background 	&		CDF	& \DZero \\
\hline
$Z\rightarrow \tau^+\tau^-$ 
  &  $0.38\pm0.07$  & $0.16\pm0.09$    \\
Drell Yan
  &  $0.44\pm0.28$  & $0.26\pm0.06$    \\
Fake $e^\pm$\ or $\mu^\pm$
  &  $0.23\pm0.15$  & $0.16\pm0.08$    \\
$W^+W^-/W^\pm Z^\circ$
  &  $0.38\pm0.07$  & $0.04\pm0.03$    \\
Heavy quarks
  &  $0.03\pm0.02$  & $0.03\pm0.03$    \\
\hline
Total 
  &  $1.3\pm0.3$  & $0.65\pm0.15$    \\
\hline
\end{tabular}
\end{center}
\caption{The number of background events expected to survive 
the CDF and \DZero\ dilepton analyses.  Only the $W^+W^-$ and heavy quark
rates are estimated based on Monte Carlo calculations in the CDF
analysis.  The other estimates are derived from background rates obtained
directly from data studies.
}
\label{tab: dilepton backgrounds}
\end{table}

The significance of this observation can be quantified in a number of
ways.  One method is to ask how likely this
observation is in the absence of 
\ttbar\ production (the null hypothesis).  
The answer to this is an
exercise in classical statistics 
\cite{ref: PDG limit calculation}, where one convolutes the
Poisson distribution of expected background events with the
uncertainty in this expected rate.  The significance of the CDF
observation is then $3\times10^{-3}$; the significance of the
\DZero\ observation is $3\times10^{-2}$.

In themselves, each analysis cannot rule out the possibility that
the observed events may be due to background sources.  Taken
together, however, they make the background-only hypothesis 
very unlikely.\footnote{
One cannot simply multiply the two significances together.  To combine
these observations, one could
define a single statistic (like the total number of observed events
in both experiments) and then model the fluctuations of this variable
in the case of the null hypothesis.  This would give a larger probability
of a background hypothesis than the product of the two probabilities.}
The obvious next step is to seek independent confirmation.

\subsection{B Tagging in the Dilepton Sample}
If the dilepton sample has a contribution from \ttbar\ production, it
is reasonable to search for evidence that two $b$\ quarks are
being produced in association with the dilepton pair and neutrinos.

The CDF collaboration has examined these events for such indications
using the $b$\ tagging algorithms described in detail in the
following section.  Three of the six events have
a total of five tagged jets, three with SLT tags and two with SVX tags.
CDF estimates that only 0.5 events with tags would be expected from non-\ttbar\
standard model sources, whereas one would expect 3.6 tags if the
events arose from the expected mixture of background and \ttbar\
production.  The data are certainly consistent with the \ttbar\ hypothesis,
and motivate a detailed study of the other potential channels.

\section{The Lepton+Jets Top Quark Search}
\begin{figure}
\vspace{2.7in}
\vskip 4.3in
\hskip -0.3in
\includegraphics{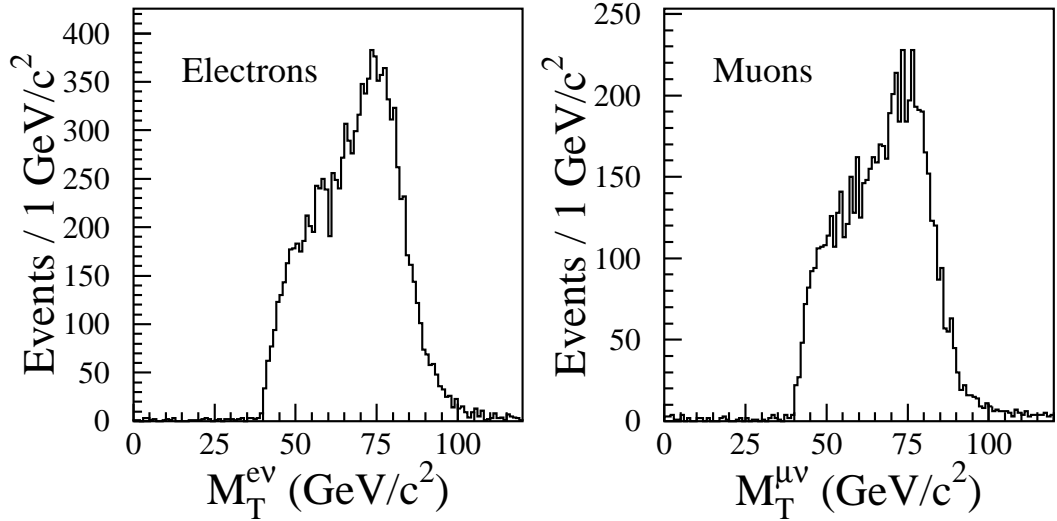}
\vskip -4.3in
\caption{The transverse mass distribution for the CDF electron and
muon samples after requiring a well-identified charged lepton and
missing transverse energy $>20$\ GeV. These data are from Run IA only.}
\label{fig: CDF Inclusive W Mt}
\end{figure}

Both collaborations begin their lepton+jets analysis from a data
sample dominated by inclusive $W^+$\ production.  They
require events with significant \MEt\ and a well-identified, high 
transverse momentum  electron or muon. 
\DZero\ requires the presence of an isolated electron with 
$E_T>20$~GeV, and \MEt $>25$~GeV to identify an inclusive 
$W^+\rightarrow e^+\nu_e$\ sample, and an isolated muon with $p_T>15$~\GeVc\
and \MEt\ $>20$~GeV to identify a $W^+ \rightarrow \mu^+ \nu_\mu$\ 
sample.
CDF requires  a candidate event to have \MEt\ $>20$~GeV and a charged
lepton candidate in the central detector with $P_T>20$\ \GeVc\ and
$|\eta|<1.0$.  The transverse mass for the resulting candidate
events, defined as
\begin{eqnarray}
M_T \equiv \sqrt{2E_T\mMEt(1-\cos\phi_{l\nu})},
\end{eqnarray}
where $\phi_{l\nu}$\ is the azimuthal opening angle between the charged
lepton and the \MEt\ vector,
has a distribution with 
a clear Jacobian peak, as illustrated by the CDF data
shown in Fig.~\ref{fig: CDF Inclusive W Mt}. 

\subsection{The \DZero\ Lepton+Jets Search}
 
\subsubsection{The \DZero\ Kinematic Analysis}
The production of $W^+$\ bosons accompanied by additional jets form
the largest single background in the lepton+jets search.  However,
there are significant differences in the kinematics of the partons in
the \ttbar\ and $W$+jets final state that can be used to
differentiate between these processes.
For example, the $H_T$\ distribution is compared for the \ttbar\ and
$W$+jets final state in Fig.~\ref{fig: HT distributions}(b).  One sees
that this variable provides significant separation between signal
and background with only a modest loss of signal.

The \DZero\ collaboration
defines a \ttbar\ candidate sample by requiring that
$H_T>200$~GeV, that
there be at least four jets in the final state with $E_T>15$~GeV and
$|\eta|<2.0$, and that the aplanarity of the event $\aplanarity>0.05$.  
This leaves 5 $e^+ +$\ jet events and
3 $\mu^+ +$\ jet events in the sample.  They expect
to observe $3.8\pm0.6$\ events from \ttbar\ production in this
sample for a top quark mass of 180~\GeVcc.

\begin{figure}
\vspace{4.5in}
\includegraphics{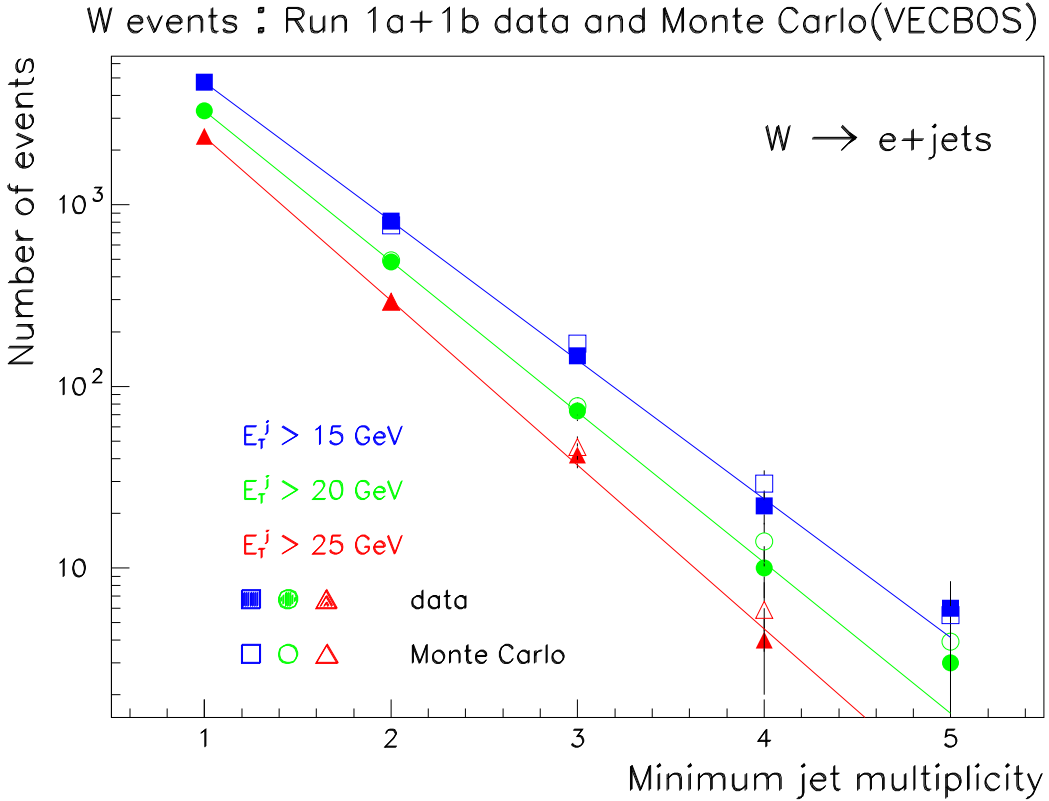}
\caption{The rate of $W^+\rightarrow e^+\nu_e$\ events as a function of the 
minimum jet multiplicity and jet $E_T$\ requirements observed by 
the \DZero\ collaboration (the charged conjugate mode is implied).  
These data are shown before the $H_T$\ or aplanarity
cuts, and are compared to predictions from a QCD Monte Carlo calculation.  
}
\label{fig: W jet Counting}
\end{figure}

The backgrounds to \ttbar\ production in this sample are dominated by
the inclusive $W$+jets process.  In order to estimate the size of
this background, one can use the rate of
observed events in the $W +1$, $W +2$, and $W +3$\ jet sample and
extrapolate that to the number of events in the $W +\ge4$\ jet sample.
It is expected that the ratio of $W +n$\ jet events to $W +(n-1)$\ jet
events will be constant given the same 
jet requirements\cite{ref: VECBOS calculations}\ when the 
$H_T$\ and aplanarity cuts are removed.
This prediction can be tested using the
$W +1$\ jets, $W +2$\ and $W +3$\ jet samples where one expects
to see little \ttbar\ contribution.  The results of this test, shown
in  Fig.~\ref{fig: W jet Counting}, confirm 
that this ratio remains constant.  

The \DZero\ collaboration then applies the $H_T$\ and aplanarity cuts and
uses the relative efficiency of these cuts on \ttbar\ signal and the
$W+$jets background to extract the number of \ttbar\ events in the sample
and the number of background events that remain.
The \DZero\ collaboration estimates the size of the background
in their $W +4$\ jet sample to be $1.9\pm0.5$\ events.
There is a clear excess of observed events above the predicted
background.

\subsubsection{B Tagging in the \DZero\ Sample}
\DZero\ has
performed a separate analysis requiring that one of the jets also be
consistent with a $b$\ quark semileptonic decay.  This study is 
complementary to the \DZero\ kinematical analysis, and does not
depend on the jet-scaling arguments to estimate the backgrounds.

\DZero's excellent muon identification capability makes it possible
to tag $b$\  hadrons by searching for the decay $b\rightarrow
\mu^-\bar{\nu}_\mu X$.  Because there are two $b$\ jets in each \ttbar\
signal event, the fraction of tagged events will be twice the
semileptonic branching fraction of $b$\ hadrons times the efficiency
for identifying muons.
\DZero\ studies show that the use of standard muon identification
requirements applied to candidates with \Pt$>4$\ \GeVc\
result in a tagging efficiency for $W +\ge3$\ jet events of $\sim
20$\%.  This is relatively insensitive to the actual top quark
mass, rising slowly as a function of $M_{top}$.

``Fake'' tags are expected to arise from real muons resulting from
heavy quark ($b$, $c$) semileptonic decay and decays-in-flight of $\pi$\ and
$K$\ mesons.  This would imply that the fake rate per jet should be
relatively independent of the number of jets in a given event, or the
topology of the jets in the event.
The \DZero\ collaboration
has measured the expected background rate for their tagging scheme
using a large sample of events coming from their inclusive jet triggers.
Since the jets in these events are expected to arise predominantly from
light quarks and gluons, they form a good sample to estimate the
probability of incorrectly $b$~tagging a light quark or gluon jet. 
This leads to an over-estimate of the
background from light quark jets, as some of the jets in this
inclusive jet control sample will have $c$\ and $b$~quarks in them, albeit
at a low rate.
These studies show that the tag rate is between 0.005 and 0.010
per jet, and rises slowly with the $E_T$\ of the jet.
Detailed Monte Carlo calculations using a full detector simulation
verify this result.
Based on this study, \DZero\ expects that $\sim2$\%\ of the $W +3$\ 
and $W +4$\ jet
background events will be tagged.
With this fake rate, $b$~tagging provides an order of magnitude
improvement in signal-to-noise in this sample.

The \DZero\ collaboration use a less stringent $W+$jets selection when
also requiring a $b$~quark tag in order to optimise the signal-to-noise of
this analysis.
The events are required to have $H_T>140$~GeV, and 
the jet multiplicity requirement is relaxed to demand at least three jets with
$E_T>20$~GeV.
In addition, the aplanarity cut is dropped altogether, and in the case
of the electron + jets channel, the \MEt\ cut is relaxed to require
$\mMEt>20$~GeV.
There are 3 events in the $e$+jet and $\mu$+jet channels that survive
these requirements, whereas only $0.85\pm0.14$\ and $0.36\pm0.08$\ events
are expected from background sources, respectively. 
As in the dilepton and lepton + jets channels, a excess of candidate events
over background is observed.

\subsection{The CDF Counting Experiment}
The CDF collaboration has performed an analysis of their
lepton+jets data similar to that reported for the Run IA
dataset
\cite{ref: CDF PRD/PRL}.  The analysis avoids making
stringent kinematical cuts that could result in large systematic
uncertainties, and takes advantage of the presence of two $b$~quarks
in the signal events to control the expected backgrounds.

Starting from the inclusive $W^+$\ boson sample, the CDF
analysis requires at least three jets with $E_T>15$\ GeV and
$|\eta|<2.0$.  This results in 203 events, with 164 and 39 events in
the $W +3$\ and $W +\ge4$\ jet samples, respectively.  
The backgrounds estimated to
make the largest contribution to this sample come from real
$W^+$\ boson production, from standard model sources of
other isolated high $E_T$\ leptons (such as $Z^\circ$\ boson production),
from $b$\ and $c$~quark semileptonic decays 
and from events where the lepton candidate has been
misidentified.  
Most of the non-$W^+$\ boson backgrounds have lower \MEt, and are
characterised by 
lepton candidates that are not well
isolated from other particles in the event.  
The correlation between this additional energy flow and \MEt\ in the event
allows one to directly measure this background fraction.
This results in an estimate for the background from sources of non-isolated
lepton candidates of
$10\pm5$\%.  
The background rates from sources that produce isolated lepton candidates 
have been estimated using data and Monte Carlo
calculations.  
These background estimates are summarised in 
Table~\ref{tab: SM W jets Backgrounds}.

\begin{table}
\begin{center}
\begin{tabular}{lc}
\hline
Background 	&		Fraction of Sample (\%)\\
\hline
$WW$, $WZ$\ Production  			&   $5.0\pm2.3$ \\
$Z^\circ\rightarrow e^+e^-/\mu^+\mu^-$\ 	&   $5.2\pm1.3$ \\
$Z^\circ\rightarrow \tau^+\tau^-$\ 		&   $3.3\pm1.0$ \\
Fake Leptons, Conversions, $b\bar{b}$\ 		&   $10.0\pm5.0$ \\
\hline
Total  						&   $23.5\pm5.7$ \\
\hline
\end{tabular}
\end{center}
\caption{The estimated fractions of events in the $W+\ge3$\ jet sample
arising from the
different background sources to \ttbar\ production. 
Only the requirement of at least three jets has been
imposed.}
\label{tab: SM W jets Backgrounds}
\end{table}

\subsubsection{Secondary Vertex Tagging}

\begin{figure}
\vspace{3in}
\hskip 0.5in
\includegraphics{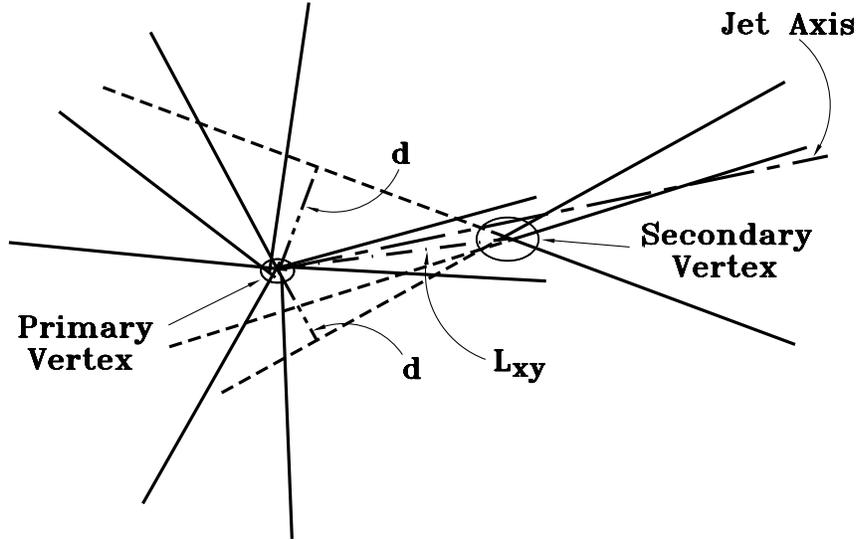}
\caption{A schematic of the decay of a bottom quark, showing the
primary and secondary vertices, and the
charged tracks reconstructed in the CDF CTC and SVX detectors.
}
\label{fig: secondary vertex}
\end{figure}

The CDF detector has the unique capability of detecting $b$~quarks by
reconstructing the location of the
$b$~quark's decay vertex using the SVX detector.  
A schematic of the
decay topology for a bottom hadron is shown in 
Fig.~\ref{fig: secondary vertex}.  The charged particle trajectories
are reconstructed in the CTC and then
extrapolated into the SVX detector to identify the track's hits in
the silicon strip detector.  

The quality of the
reconstructed SVX track is determined by the number of SVX coordinates
found for the track and the accuracy of each coordinate. 
The algorithm to reconstruct secondary vertices
considers all tracks above a transverse momentum of 1.5 \GeVc\ that
have an impact parameter relative to the primary vertex $>2\sigma$,
where $\sigma$\ is the estimated uncertainty in the impact parameter
measurement for the track.
The algorithm first looks for vertices formed by three tracks,
making relatively loose quality cuts on each of the tracks.   A vertex
is accepted if a $\chi^2$\ fit requiring the three tracks to come from
a common point is acceptable.   Any remaining high-quality tracks with
large impact parameter are then paired up to look for two-track
vertices.  A jet containing a secondary vertex found in this way that
has a positive decay length is considered SVX tagged (the sign of the
decay length is taken from the dot product of the  displacement vector
between the primary and secondary vertices, 
shown as $L_{xy}$\ in Fig.~\ref{fig: secondary vertex},
and the vector sum of the
momenta of the daughter tracks).

The efficiency of this SVX tagging algorithm has been measured using
a large sample of inclusive electron and 
$J/\psi\rightarrow\mu^+\mu^-$\ candidates,  where the heavy quark 
contents in these samples have been independently estimated.  This
efficiency agrees with that obtained using a full detector
simulation;  the ratio of the measured efficiency to the efficiency
determined using the detector simulation
is $0.96\pm0.07$.

The $b$~quark SVX tags not arising from \ttbar\ production 
come from track combinations that for some reason
result in a fake secondary vertex (mistags) and from real sources of $b$\ and 
$c$~quarks in $W +$\ jet events.
One way of estimating the mistag rate is to
note that the rate of these fakes must be equal 
for those secondary vertices located on either side of the \pbarp\  
collision vertex as determined by comparing the displacement vector of
the secondary vertex with the momentum vector of the tracks defining the
secondary vertex
(positive and
negative tags, respectively).
The rate of real $b$\ and $c$~quarks not arising from \ttbar\ production
can be estimated using theoretical calculations and comparing these with
observed rates in other channels.

\begin{figure}
\vspace{4.8in}
\vskip 1.3in
\hskip 0.0in
\includegraphics{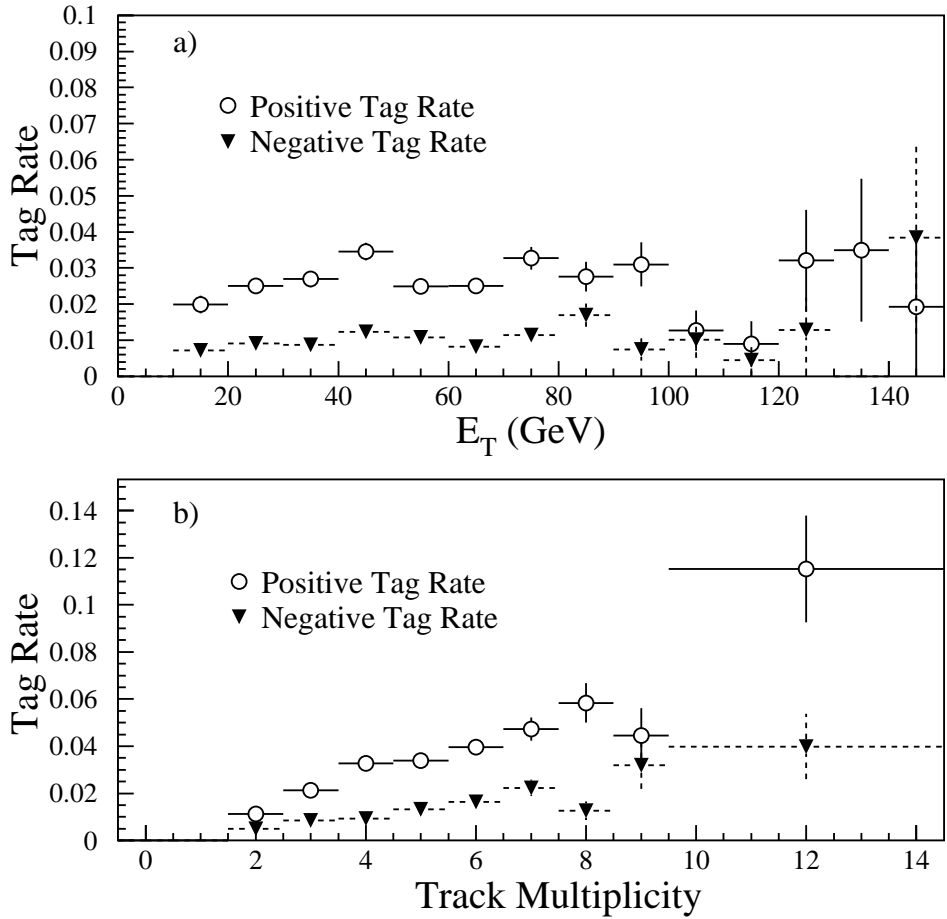}
\vskip -1.3in
\caption{The rate of SVX tags as a function of the transverse
energy of the jet and the charged track multiplicity in the jet,
as measured using the inclusive jet sample.  Tag rates for both positive and
negative decay length vertices are shown.}
\label{fig: SVX mistag rate}
\end{figure}

The mistag probability has been measured using both samples of
inclusive jets and the inclusive electron and dimuon samples.  The
probability of mistagging as a function of the number of jets in the
event and the transverse energy of
the jet is shown in Fig.~\ref{fig: SVX mistag rate}, based on the
inclusive jet measurements where I have plotted both the negative and positive
tag rates.
The negative tag rate is perhaps the best estimate of the mistag rate,
since one expects some number of real heavy quark decays in
this sample to enhance the positive tag rate. 
The mistag rate per jet measured in
this way is $\sim0.008$, and is lower than the positive tag rate measured
in the inclusive jet sample ($\sim0.025$), as expected from estimates of
heavy quark production in the inclusive jet sample.

To account for all sources of background tags, the number of
tagged events expected from sources of real heavy quark decays 
(primarily $W^+b\bar{b}$\ and $W^+c\bar{c}$\ final states) is 
determined using a Monte Carlo calculation and a full simulation of the
detector.  
The sum of this ``physics'' tag rate and the mistag rate then gives an
estimate of the total background to \ttbar\ production.
This estimate can be checked by using the positive tag rate in inclusive
jet events as a measure of the total non-\ttbar\ tag rate in the
$W +$\ jet events.  This gives us a somewhat higher background rate, 
due primarily to the expected larger fraction of $b$\ and $c$~quarks in the
inclusive jet sample compared to the $W+$jet events.

The efficiency for finding at least one jet with an SVX tag in a 
\ttbar\ signal event is calculated using the ISAJET Monte Carlo 
programme \cite{ref: ISAJET}\ to
generate a \ttbar\ event, and then applying the measured tagging
efficiencies as a function of jet \Et\ to determine how many reconstructed 
$b$~quark jets are tagged.  The SVX tagging efficiency, \ie\ the fraction of 
$W+\ge3$~jet \ttbar\ events with at least one SVX-tagged jet, 
is found to be $0.42\pm0.05$,
making this technique a powerful way of identifying \ttbar\ 
candidate events.

\subsubsection{Soft Lepton Tagging}
The CDF collaboration developed the original lepton-tagging
techniques to search for $b$~quarks in \ttbar\ production 
\cite{ref: CDF 88-89 Top PRD}, requiring the presence of a 
muon candidate in proximity to one of the jets.  
The collaboration has enhanced these techniques by extending the
acceptance of the muon system and by searching for
electron candidates associated with 
a jet cluster.  In both cases, it is optimal to allow for
relatively low energy leptons (down to \Pt's as low as 2 \GeVc), so
this technique has become known as ``soft lepton tagging.''
A candidate jet cluster with a soft lepton candidate is
considered to be SLT tagged.  

The efficiency of this tagging technique depends on the
ability to identify leptons in the presence of additional hadrons
that come from the fragmentation of the $b$~quark and the decay of
the resulting $c$~quark system.  Muons are identified by requiring
a charged track in the CTC that matches a muon track stub.  Electron
candidates are defined by an electromagnetic shower in the
calorimeter with less than 10\%\ additional energy in the hadronic
calorimeter towers directly behind the shower, a well-reconstructed
track in the CTC that matches the position of the shower and shower
profiles consistent with those created by an electron.
The overall efficiency for finding at least one SLT tag in a \ttbar\
event is $0.22\pm0.02$, and is not a strong function of the top quark
mass.  

The rate at which this algorithm misidentifies light quark or gluon
jets as having a soft lepton is determined empirically by studying 
events collected by requiring the presence of at least one
jet cluster.  The mistag rate for muon tags varies between 0.005 and
0.01 per charged track, and
rises slowly with the energy of the jet.  The mistag
rate for electrons also depends on the track momentum and how
well isolated it is from other charged tracks; it typically is of
order 0.005 per track.
Fake SLT tags where
there is no heavy flavour semileptonic decay is expected to be the
dominant source of background tags in the \ttbar\ sample, due to the larger
SLT fake rates as compared to the SVX mistag rates.  

\subsubsection{Tagging Results in the CDF Lepton+Jets Sample}
The SVX and SLT tagging techniques have been applied to the 
$W+$jet sample as a function of the number of jets in the event,
and the expected number of mistags has been calculated for each
sample.
This provides a very strong consistency check, as the number of
observed tags in the $W +1$\ jet and $W +2$\ jet samples should be 
dominated by background tags; the fraction 
in these two event classes expected from \ttbar\ production is less
than 10\%\ of the total number of candidate events.

The number of candidate
events and tags is shown in Table~\ref{tab: lepton+jet B tags}.  There
is good agreement between the expected number of background tags and the
number of observed tags for the $W +1$\ jet and $W+2$\ jet samples. 
However, there is a clear excess of tags observed in the $W+\ge3$\ jet
sample, where we observe 27 and 23 SVX and SLT events, respectively,
and expect only
$6.7\pm2.1$\ and $15.4\pm2.3$\ SVX and SLT background tags.  The excess of
SVX tags is particularly significant, with the probability of
at least this number of tags arising from background sources being
$2\times10^{-5}$.  The excess of SLT tags is less significant
because of the larger expected background.  The
probability that at least 23 observed SLT tags would arise from background only
is $6\times10^{-2}$\ and confirms the SVX observation.

\begin{table}
\begin{center}
\begin{tabular}{ccccc}
\hline
Sample     & SVX bkg   &  SVX tags  &  SLT bkg &  SLT tags \\
\hline 
W+1 jet    & $50\pm12$  &  40     & $159\pm25$ &  163   \\
W+2 jet    & $21\pm7$   &  34     & $46\pm7$   &  55   \\
W+$\ge3$ jet & $6.7\pm2.1$ &  27   & $15.4\pm2.3$   &  23  \\
\hline
\end{tabular}
\end{center}
\caption{The expected number of background tags and the observed number
of tags in the CDF lepton+jets sample as a function of the number
of jets in event. }
\label{tab: lepton+jet B tags}
\end{table}

It is interesting to note that if we attribute the excess number of
SVX tags in the $W +\ge3$\ jet sample to \ttbar\ production, we would
expect approximately 10 $W+2$\ jet tagged events resulting from
\ttbar\ production.  This is in good agreement with the excess of
observed tags ($13\pm7$) in this sample, and corroborates the
hypothesis that the excess in the $W +\ge3$\ jet sample is due
to the
\ttbar\ process.

\begin{figure}
\vspace*{6.0in}
\vskip 0.0in
\includegraphics{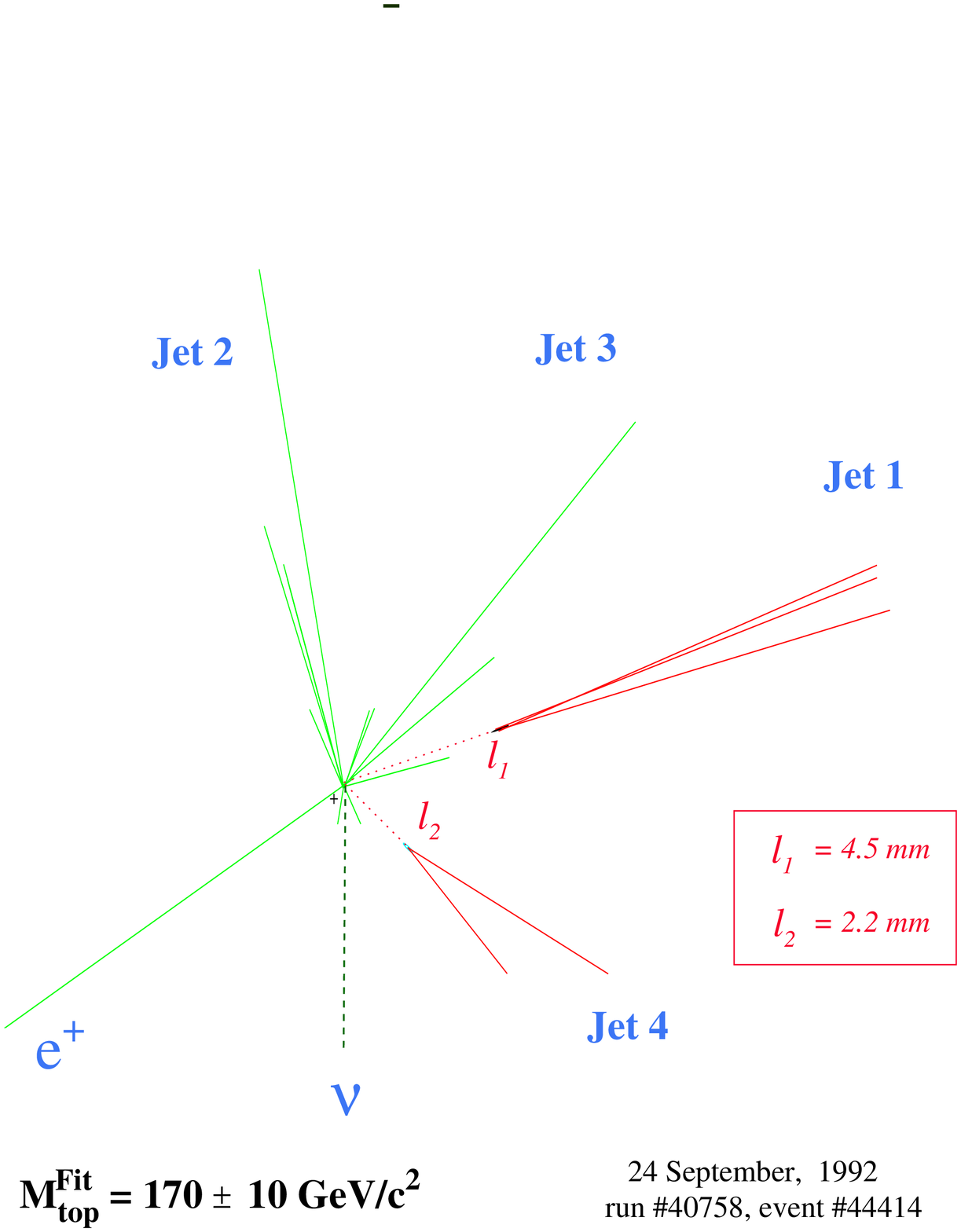}
\vskip -0.0in
\caption{
The schematic in the $r$-$\phi$\ view 
of the SVX tracks reconstructed in one of the
CDF lepton+jet events that has two SVX tagged $b$~jets.
The jets associated with the SVX tracks
and the lepton candidates are identified.  The decay lengths of
each $b$~candidate jet are noted in the figure.
This event is fitted to a top quark mass of $170\pm10$~\GeVcc, using
the procedure discussed below.
}
\label{fig: SVX double tag}
\end{figure}

A striking feature of the tagged sample is the number of events with
two or more tagged jets.  The 27 SVX tags are found in 21 events, so
that there are 6 SVX double tags.  There are also six SVX tagged
events that have SLT tags.  We would expect less than one SVX-SVX
double tag and one SVX-SLT double tag in the absence of
\ttbar\ production, whereas we would expect four events in each
category using the excess of SVX tags to estimate the \ttbar\
production cross section.  
A schematic of one of the SVX double tagged events is shown in 
Fig.~\ref{fig: SVX double tag}, where the tracks reconstructed in the
SVX detector are displayed, along with the jets and lepton candidates they
are associated with.
These observations strengthen the \ttbar\
interpretation of the  CDF sample.

\subsection{Summary of Counting Experiments}

The results of the lepton+jets counting experiments performed by
\DZero\ and CDF are summarised in Table~\ref{tag: counting summary}.
Both collaborations observe an excess of events in all the channels
in which one can reasonably expect evidence for the top quark.  Many of
the channels demonstrate correlated production of $W^+$\ bosons
with $b$~quarks -- exactly what we would expect from \ttbar\ decay.

\begin{table}
\begin{center}
\begin{tabular}{lcc}
\hline
Sample				&   Background & Observed \\
\hline
CDF Dileptons   		& $1.3\pm0.3$ 	&   6   \\
\DZero\ Dileptons   		& $0.65\pm0.15$	&   3   \\
Lepton + Jets (\DZero\ Kinematics)    & $0.93\pm0.50$   & 8 \\
Lepton + Jets (\DZero\ B Tagging)     & $1.21\pm0.26$   & 6 \\
Lepton + Jets (CDF SVX tags)    &   $6.7\pm2.1$ &  27  \\
Lepton + Jets (CDF SLT tags)    &  $15.4\pm2.3$ &  23  \\
\hline
\end{tabular}
\end{center}
\caption{The expected number of background events and the observed
number of events in the different analyses.  Note that some event
samples and background uncertainties are correlated so it is not
straightforward to combine these observations into a single 
statement of statistical significance.  }
\label{tag: counting summary}
\end{table}

Taken together, this is overwhelming evidence that the
two collaborations are observing phenomena that within the context of
the standard model can only be attributed to pair production of top
quarks.  

\section{Measurement of Top Quark Properties}
In order to further test the interpretation that top quark
production is responsible for the excess in the dilepton and
lepton+jets channels, both
collaborations have measured the rate of top quark production and
identified a subset of their candidate lepton+jet events where it is
possible to directly measure the mass of the top quark. 

These measurements allow us to test the 
standard model prediction for the cross section as a
function of the top quark mass.  The initial 
evidence for top quark production published by CDF 
\cite{ref: CDF PRD/PRL}\ implied a top
quark production cross section almost two standard deviations above
the theoretically predicted value.  Moreover, other standard model
measurements, and in particular those performed at LEP, constrain the
top quark mass.  It is important to directly verify that
these predictions agree with the top quark mass inferred from the
Collider data. 

The CDF and \DZero\ Collaborations have also begun other studies of top quark
properties that can be inferred from the Collider data.  These include
aspects of both top quark decay and production, and I discuss their 
status in the following subsections.

\subsection{The Top Quark Cross Section}
The acceptance of the \DZero\ and CDF top quark searches depend on
the top quark mass.  We can therefore infer the \ttbar\ production
cross section as a function of the top quark mass given the number of
observed events in each channel.

For a data sample with integrated luminosity ${\cal L}$,
if we observe $N^o_i$\ candidate
events in a particular channel $i$\ 
and we expect $N^b_i$\ background events, then the 
maximum likelihood solution for the cross section of the process 
combining all channels is
\begin{eqnarray}
\sigma = {{\sum_i \left(N^o_i-N^b_i\right)}\over{{\cal L}\left(\sum_i
\epsilon_i\right)}},
\end{eqnarray}
where $\epsilon_i$\ is the acceptance for the search.
This assumes that the observed number of events has a Poisson
distribution and that uncertainties on the acceptance can be
ignored. The latter restriction can be relaxed by numerically solving
for the maximum likelihood solution allowing for uncertainties 
in $\epsilon_i$\ and $N^b_i$, and any
correlations in the acceptances. 

The CDF collaboration has performed a preliminary measurement of the
\ttbar\ cross section using the SVX tagged sample.  This is
the single most significant measurement and can be performed only
knowing the SVX tagging efficiency and background rates.  The addition
of the SLT sample and the dileptons into the cross section
measurement requires a knowledge of the efficiency correlations in
the samples and is work in progress.    The
\ttbar\ acceptance was determined using the ISAJET Monte Carlo
programme, and found to be $0.034\pm0.009$.  The uncertainties
associated with this acceptance calculation are listed in 
Table~\ref{tab: acceptance uncertainties}.
The expected background in the 21 tagged events is $N^b=5.5\pm1.8$\ 
events.\footnote{
The previous estimate of the expected SVX background tags assumed that there
was no contribution from \ttbar\ production to
the 203 events in the $W+\ge3$\ jet sample prior to tagging.} 

\begin{table}
\begin{center}
\begin{tabular}{lc}
\hline
Source   &   Uncertainty (\%) \\
\hline
Lepton ID and Trigger	& 10 \\
Initial State Radiation &  7 \\
Jet Energy Scale        & 6.5 \\
$b$~Tagging Efficiency	& 12 \\
\hline
\end{tabular}
\end{center}
\caption{The uncertainties in the acceptance calculation for the
CDF cross section measurement using the SVX tagged sample.}
\label{tab: acceptance uncertainties}
\end{table}

The resulting cross section determined from the SVX sample is 
$6.8^{+3.6}_{-2.4}$\ pb\ for a nominal top quark mass of
175 \GeVcc. This is approximately one standard deviation lower than the
cross section determined in the Run IA CDF data.  It is in good
agreement with the theoretically predicted value of $4.9\pm0.6$\ pb for
the same top quark mass.

\begin{figure}
\vspace{3in}
\vskip 0.3in
\hskip 0.5in
\includegraphics{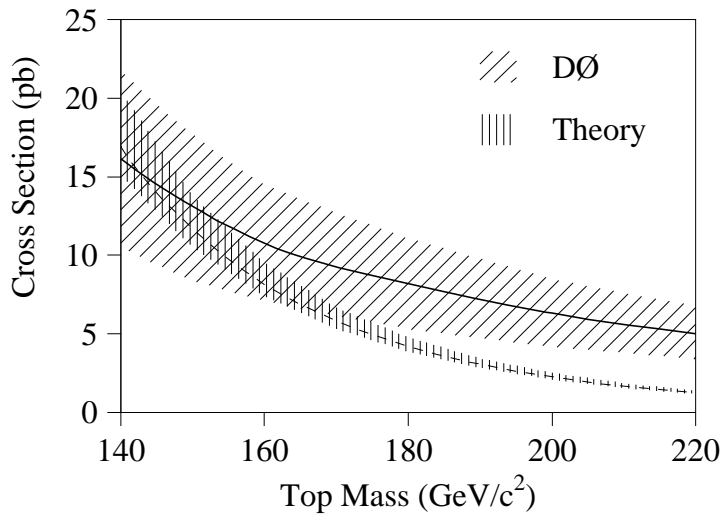}
\vskip -0.3in
\caption{The top quark cross section determined by the \DZero\ 
collaboration as a function of top quark mass.  The QCD prediction
for \ttbar\ production is displayed as the heavier band.
}
\label{fig: D0 cross section}
\end{figure}

The \DZero\ collaboration estimates the \ttbar\ cross section using
the information from all the channels they have studied.  
They also perform a background
subtraction and then correct for the acceptance, channel by channel.
They determine $\sigma_{\mttbar} = 6.2\pm 2.2$~pb, for a top quark mass
of 200~\GeVcc.  This value doubles to $\sim12$~pb if one
assumes a top quark mass of 160~\GeVcc.
The top quark mass dependence of the \DZero\ cross section is illustrated in
Fig.~\ref{fig: D0 cross section}.

The CDF and \DZero\ estimates are in reasonable agreement with each other, 
although both have large uncertainties.  A 
strong test of the lowest order calculation for $\sigma_{\mttbar}$\ 
and next-to-leading order corrections will have to wait for substantially
more statistics.

\subsection{The Top Quark Mass}
The top quark mass can be determined directly by correlating the
kinematics of the observed partons in the final state.  The
sensitivity of this measurement depends on the amount of ``missing''
information in the events, and the inherent resolution of the
detectors to jets and missing energy.  The lepton $+\ge4$~jet events offer
the possibility of fully reconstructing the \ttbar\ system provided
one assumes that the missing transverse energy arises from the
undetected neutrino, and that four of the jets come from the $b$\ and $\bar{b}$\
quarks and the two quarks from the $W^+$\ decay.

Perhaps the most serious complication to this procedure is the difficulty of
associating final state jet clusters with the partons from the
\ttbar\ decay.  The jets are only approximate measures of the initial
state parton, and there is often not a 1-to-1 correspondence between
partons resulting from the \ttbar\ decay and observed jets.
This is due to 
gluon radiation that can cause one parton to be observed as two jet
clusters, and overlap of jet clusters, where two partons merge
into a single jet cluster.  To complicate matters further, additional
partons are produced by initial and final state radiation, so 
the number of observed jet clusters may readily exceed four. 

The number of combinatorial possibilities for assigning partons to
jets in the case where only four jets are observed is twelve (we only have
to identify the two jets associated with the $W^+$\ decay and not
have to permute these two).  If we can identify one of the jets as
arising from a bottom quark, the number of possible assignments
reduces to six.  
Any technique that reconstructs the \ttbar\ decay in this mode has
to reduce the effect of these combinatorial backgrounds on the
expected signal.

\subsubsection{CDF Mass Analysis}
The CDF collaboration measures the top quark mass by selecting a
sample of lepton+jet events with at least four jets, and then making
the parton-jet assignment that best satisfies a constrained kinematic
fit. The fit inputs are the observed jet momentum
vectors, the momentum vector for the charged
lepton, the transverse energy vector for the neutrino and the vector
sum of the momentum of the unassigned jets in the event.  The uncertainties
in these quantities are determined from the measured response of the detector.
The fit
assumes that the event arises from the process 
\begin{eqnarray}
p\bar{p} &\rightarrow& t\bar{t} X, \\
   && \mrightdownarrow q \bar{q}^\prime\, \bar{b} \nonumber \\
   && \mrightdownarrow l^+ \nu_l\, b \nonumber
\end{eqnarray}
The fit 
constrains the $W^+$\  and $W^-$\ decay daughters to have an invariant mass
equal to the $W^+$\ mass and constrains the $t$\ and the $\bar{t}$\ to
have the same mass.  The unknown recoil system $X$\ is observed in the 
detector as unassociated jets and the ``unclustered'' energy in the
calorimeter, \ie\ the energy not associated with a jet.  
Only the four highest \Et\ jets are considered, reducing the possible
combinations at the cost of some degradation in top quark mass resolution
(in those cases where the \ttbar\ daughter jets are not the four highest
\Et\ jets in the event).
 
Formally, there
are two degrees of freedom in the fit when we take into account the
number of constraints and the number of unmeasured quantities.  
A $\chi^2$\ function including
the uncertainties in the measurements is minimised subject to the
kinematic constraints for each possible parton-jet assignment.  The
$b$-tagged jets in the event are only allowed to be assigned to the $b$\
or
$\bar{b}$\ quarks.  Prior to the fit all jet energies are corrected in
order to account for detector inhomogeneities and the effect of
energy flow into and out of the jet clustering cone.  The parton
assignment that produces the lowest
$\chi^2$\ is selected for the subsequent analysis. 
The event is rejected if the minimum $\chi^2$\ is greater than 10. 
Parton assignments that result in a top quark mass greater than 260~\GeVcc\ are
also rejected as the experiment is not expected to have any sensitivity to
top quark masses of that magnitude.

\begin{figure}
\vspace{4.2in}
\vskip 1.2in
\hskip 0.5in
\includegraphics{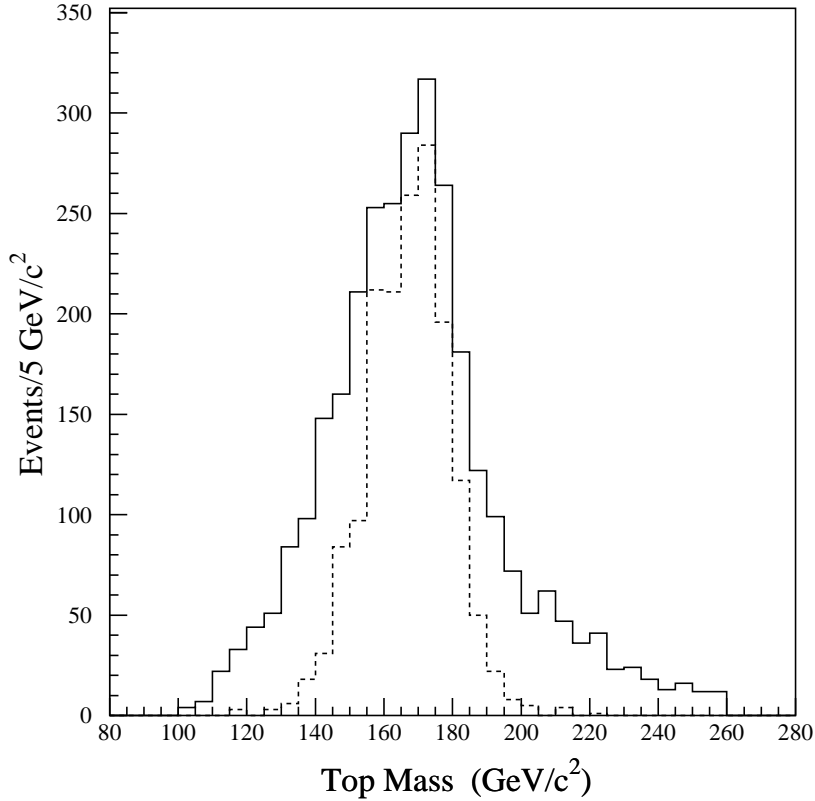}
\vskip -1in
\caption{The fitted top quark mass in Monte Carlo events for those
events in which the correct parton assignments have been made
(dashed histogram) and for all events that pass the fit procedure
(solid histogram).  A top quark mass of 170 \GeVcc\ has been assumed.}
\label{fig: MC fitted mass}
\end{figure}

Monte Carlo studies have demonstrated that this procedure identifies the
correct parton-jet assignment about 40\%\ of the time.  The top quark
mass resulting from the fit in those cases is shown in 
Fig.~\ref{fig: MC fitted mass}\ along with the mass distribution for
all lepton + $\ge4$\ jet events for a sample created assuming a top quark mass of 
170~\GeVcc.  From a single event, one is able to measure the top quark mass
to an accuracy of $\sim10$\ \GeVcc\ when one makes the
correct assignment.  However, the full distribution shows that
the fitting and parton
assignment procedure retains much of this mass information even in
those cases where the incorrect parton assignment has been made.  

Starting  with the 203 $W +\ge3$\
jet events, the CDF collaboration selects a subset of events that have
at least one additional jet with
$E_T>8$\ GeV and
$|\eta|<2.4$.  The requirements on the fourth jet are less stringent
than the first three jets in order to enhance the efficiency for
detecting all four jets from the \ttbar\ decay.  There are 99 such
events in the CDF sample prior to requiring a $b$-tagged jet, and 88 of
these pass the
$\chi^2$\ cut on the best jet-parton assignment and kinematic fit.  The
additional requirement of at least one SVX or SLT-tagged jet leaves 19
events.  

The background of non-\ttbar\ events in this sample is estimated in
the same manner used in the cross section analysis.  One 
assumes that the 88 event sample is a mixture of background and
\ttbar\ signal, and then applies the known background tag rates to determine
how many of the non-\ttbar\ events would be tagged.  This results in 
a estimated background in the 19 events of $6.9^{+2.5}_{-1.9}$\
events. This background is expected to be a combination of real $W$+jet
events and events where an energetic hadron fakes the lepton
signature.  Studies of the $Z$+jet events, candidate events where
the lepton is not well-isolated and $W$+jet Monte Carlo events
show that the resulting top quark mass distribution for these
different background events are all similar.   The CDF collaboration
therefore uses the $W$+jet Monte Carlo sample to
estimate the background shape in the top quark mass distribution.

\begin{figure}
\vspace{4.5in}
\vskip 1in
\hskip 0.5in
\includegraphics{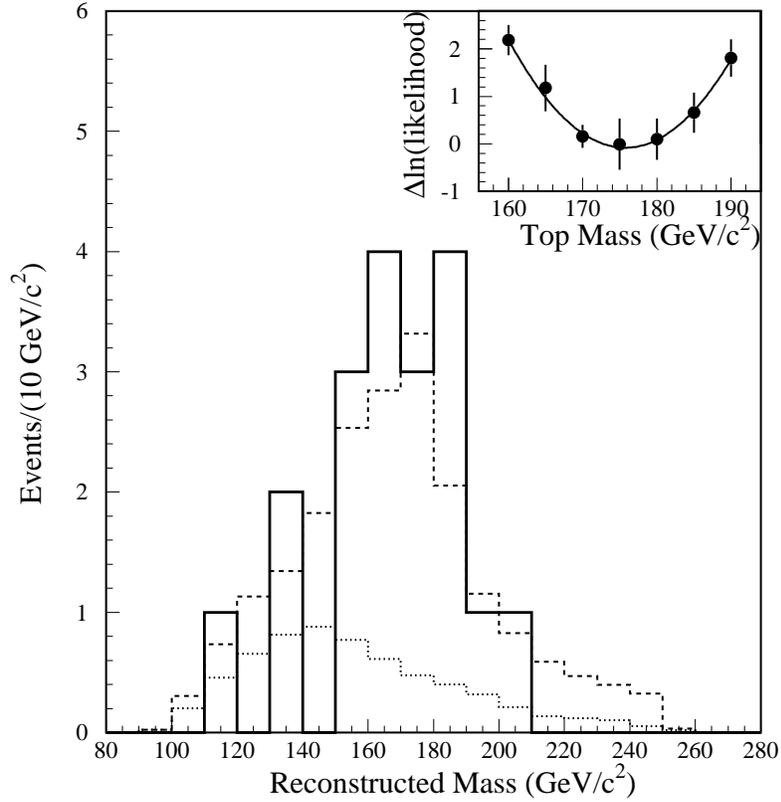}
\vskip -1in
\caption{The fitted top quark mass for the 19 events in the
CDF sample with four or more jets that satisfy the fit criteria.
The dotted histogram reflects the shape and size of the estimated
background. The dashed histogram is the result of a fit of the 
reconstructed mass distribution to a combination of \ttbar\ signal
and expected background.
The inset distribution is the change in log-likelihood of this fit.
}
\label{fig: 19 events fitted}
\end{figure}

The resulting top quark mass distribution is shown in 
Fig.~\ref{fig: 19 events fitted}.  One sees a clear peak around 170-180
\GeVcc\ with relatively long tails.  The dotted distribution
represents the shape of the non-\ttbar\ backgrounds, normalised to
the estimated background rate. The top quark mass is determined by
performing a maximum likelihood fit of this distribution to a linear
combination of the expected \ttbar\ signal shape determined by Monte
Carlo calculations for different top quark masses and the background.
The background rate is constrained by the measured rate of non-\ttbar\
events in the sample.  The negative log-likelihood distribution for
this fit is shown in the inset in Fig.~\ref{fig: 19 events fitted}.
It results in a top quark mass of $176\pm8$\ \GeVcc.

\begin{figure}
\vspace{4.5in}
\vskip 1in
\hskip 0.5in
\includegraphics{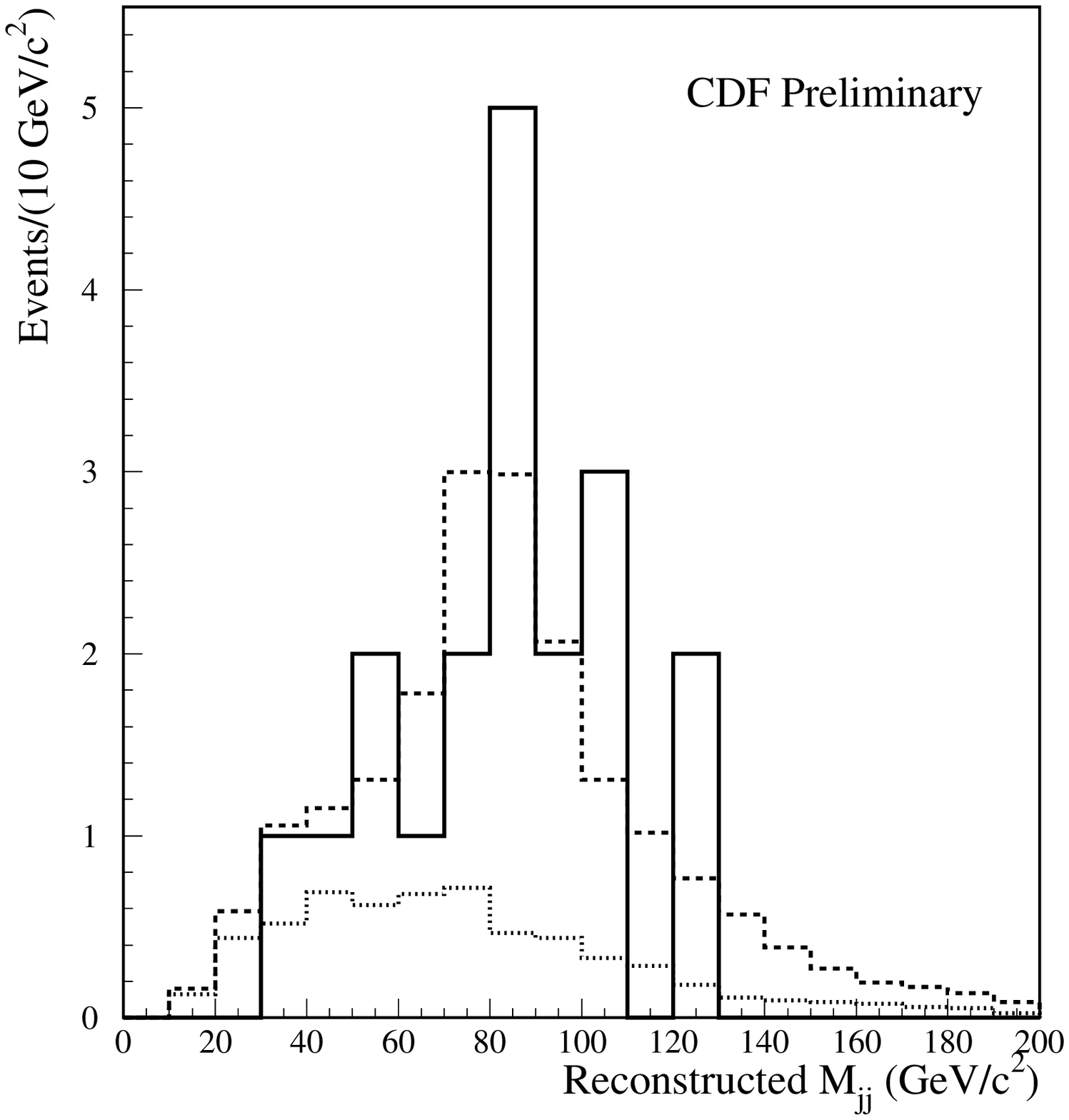}
\vskip -1in
\caption{The solid histogram is the fitted dijet invariant mass 
distribution for the $W+\ge4$\ jet events in the
CDF sample that satisfy the fit criteria.
In this case, the dijet invariant mass constraint has been relaxed and
the lowest $\chi^2$\ solution has been plotted.
The heavy dashed histogram is the expected distribution from
a combination of \ttbar\ signal and the non-\ttbar\ background.
The light dashed histogram is the background distribution 
normalised to the expected number of background events in this sample.
}
\label{fig: ttbar dijet mass}
\end{figure}

Since the fit constrains the invariant mass of the jets assigned to be
the $W^+$\ boson daughters to the $W^+$\ boson mass, one can only test
the consistency of this assignment by first relaxing this constraint and
then examining the dijet invariant mass distribution.
I show this in Fig.~\ref{fig: ttbar dijet mass}\ for the $W+\ge4$\ jet
events that satisfy the selection criteria without the imposition of 
the dijet mass constraint.  The comparison with the expected distribution
from the combination of background events and \ttbar\ signal is quite
good.  However, one should keep in mind the rather low statistics and
the large expected mass resolution.  This distribution will become a very
important calibration tool when larger statistics samples become available.

The largest systematic uncertainties in this measurement arise from
uncertainties in the modelling of gluon radiation in jets in the final
state, absolute jet energy scale, variations in fitting procedures, and the shape
of the non-\ttbar\ background.   A number of other potential sources of
uncertainty have been studied, and have been found to contribute a total
of $\pm2.0$~\GeVcc\ to the total systematic uncertainty. 
A summary of these uncertainties is given in 
Table~\ref{tab: top mass uncertainties}, and total to $\pm10$\ \GeVcc.

\begin{table}
\begin{center}
\begin{tabular}{lc}
\hline
Source   &   Uncertainty (\GeVcc) \\
\hline
Final State Gluon Radiation 		&    7.7  \\
Absolute Jet Energy Scale		&    3.1  \\
Variations in Fit Procedures		&    2.5  \\
Shifts Resulting from Tagging Biases	&    2.4  \\
Monte Carlo Statistics			&    3.1  \\
Non-\ttbar\ Mass Distribution Shape 	&    1.6 \\
Miscellaneous Effects			&    2.0 \\
\hline
\end{tabular}
\end{center}
\caption{The systematic uncertainties associated with the CDF top quark mass
measurement.}
\label{tab: top mass uncertainties}
\end{table}

One can quantify the significance of the shape of the mass
distribution by performing an unbinned Kolmogorov-Smirnov test.  The
probability that the observed mass distribution could arise from
purely background sources is $2\times10^{-2}$.  This test is
conservative in that it only compares the shape of the background
with the observed data.  Other measures of significance can be used. 
For example, one can define a relative likelihood for the
top+background and background-only hypotheses and then ask how
often a background-only hypothesis would result in a relative
likelihood as significant as that observed.  This test gives a
probability for a background fluctuation of less than $10^{-3}$.
However, it is more model-dependent as it assumes a specific shape
for the non-background hypothesis.

\subsubsection{The \DZero\ Mass Measurement}
The \DZero\ collaboration estimates the top quark mass using their
sample of lepton + $\ge4$\ jet events.  In their analysis, they select
4-jet events by requiring that all jets have a corrected transverse
energy $>15$\ GeV with $|\eta|<2.4$.  They also require the events to
have $H_T>200$\ GeV and to have aplanarity $>0.05$.
They find 14 events
that satisfy these requirements. 

\begin{figure}
\vspace{5.5in}
\vskip 0.7in
\includegraphics{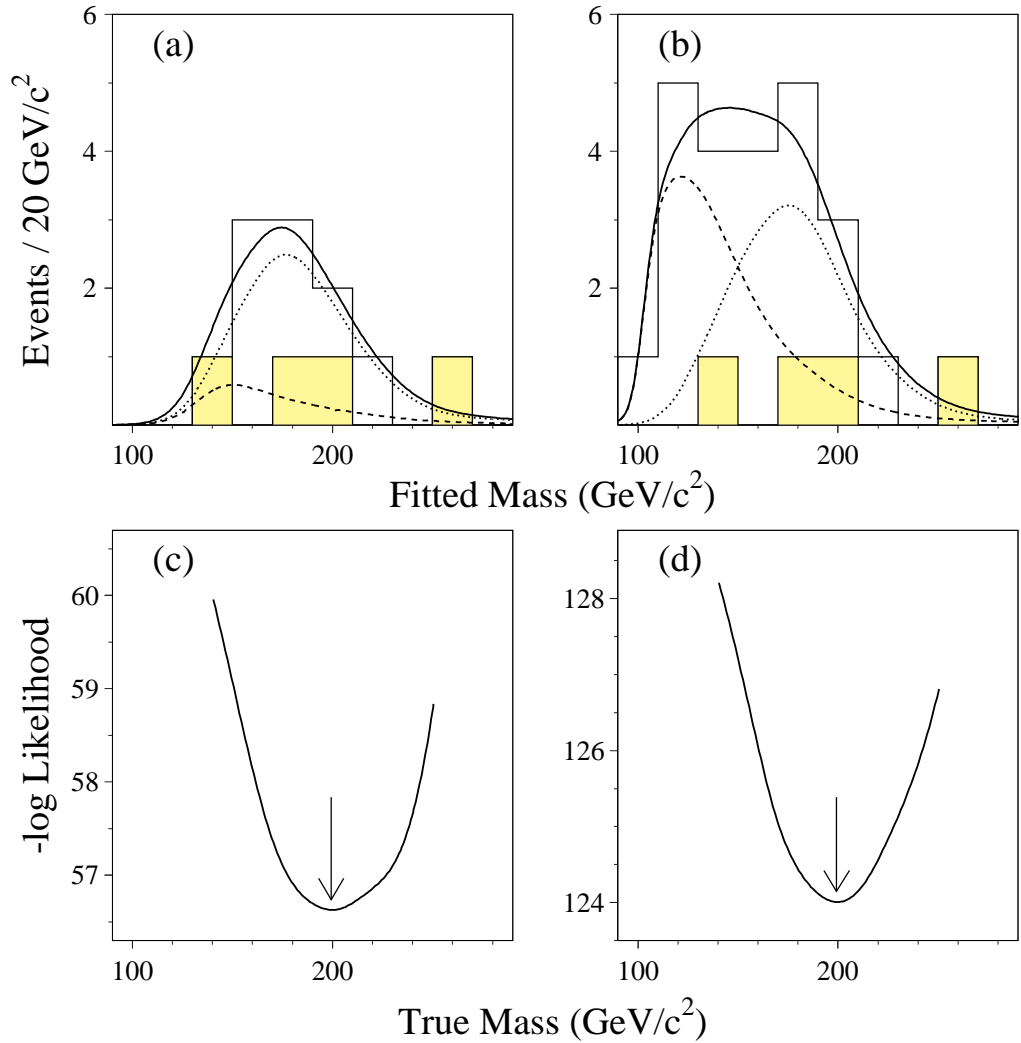}
\vskip -0.7in
\caption{The distribution of the three-jet invariant mass
versus the top quark mass obtained from the \DZero\ lepton + 4 jet
sample.
Figures a) and b) show the results of the standard and ``loose'' selection,
respectively.  
The dashed curves are the predicted background distributions, the dotted
curves are the \ttbar\ signal distributions and the solid curves are the
sum of these.
Figures c) and d)  show the likelihood distribution for fits of
the mass distributions to a combination of signal and background terms.
}
\label{fig: D0 2-D Mass}
\end{figure}

They then perform a $\chi^2$\ fit of the observed kinematics in
each event to the $t\bar{t}\rightarrow W^+ W^- b\bar{b}$\ hypothesis,
requiring that the mass of the assumed $t\rightarrow l^+\nu_l b$\ system
equal the mass of the $t\rightarrow q\bar{q}^\prime b$\ system making all
possible parton-jet assignments in the final state.  
As in the CDF technique, they only consider the four highest \Et\ jets,
and only fits with $\chi^2<7$\ are considered acceptable.
There are 11 events
that have at least one configuration that gives an acceptable fit.
For each event, they assign a top quark mass by averaging the top quark
mass from the three best acceptable fits for that event, weighting 
the mass from each fit with the $\chi^2$\ probability from the fit.
The resulting histogram of the invariant mass of the 
three-parton final state (the hypothesised top quark) is shown in
Fig.~\ref{fig: D0 2-D Mass}(a).  They performed the same analysis on a 
``looser'' data sample of 27 events, where the $H_T$\ and aplanarity requirements
were removed. This yields similar results, as 
shown in Fig.~\ref{fig: D0 2-D Mass}(b), although with significantly
larger backgrounds.
The mass distribution shows an enhancement 
at a three-parton invariant mass around 200
\GeVcc, as expected from \ttbar\ production (shown as
the higher mass curve in both plots).  
The corresponding mass distribution expected from the QCD
$W+$jet background is shown in Fig.~\ref{fig: D0 2-D Mass}(a)-(b)\ as the dashed
curve at lower mass.  It
peaks at small values of three-parton invariant mass and together the combined
background and signal hypothesis model the data well.

The mass distribution obtained using the looser selection is
fit to a combination of \ttbar\ signal and 
background, yielding a top quark mass of 
\begin{eqnarray}
M_{\rm top} = 199^{+19}_{-21} \pm 22\ \mGeVcc,
\end{eqnarray}
where the two uncertainties are statistical and systematic, respectively.
A similar fit to the mass distribution using the 11 event sample results in
a consistent result, but with larger statistical uncertainties.
The negative log-likelihood distributions for the fits to the standard and
loose selection are shown in Fig.~\ref{fig: D0 2-D Mass}(c) and (d), respectively.
The systematic uncertainty is dominated by the sensitivity of this analysis
to the \DZero\ jet energy scale.

\subsection{Top Quark Decays}
The standard model predicts that the top quark will decay via a $V$-$A$\
interaction into the $W^+ b$\ final state 100\%\ of the time.  
It is important to confirm this prediction as various extensions
to the standard model differ on the predicted phenomenology of top quark
decays.
There are
effectively two separate predictions that should be tested:
\begin{enumerate}
\item The decay proceeds via the standard model charged current. 
\item The top quark always decays to a $b$~quark.
\end{enumerate}
It is useful to address these two predictions separately as they
involve different aspects of the standard model,
namely the assumption that there is only one current involved in the
top quark decay and on our understanding of the $t W^+ b$\ vertex.

In the context of the standard model, the GIM mechanism is responsible
for suppressing all flavour-changing neutral currents (FCNC).  This has been
experimentally tested in the strange and bottom quark sector, where limits
on FCNC decays are quite stringent.\cite{ref: FCNC limits}\  
An extension to the top quark
sector is therefore an important verification of this fundamental aspect
of the electroweak interaction.  
The standard model does allow top quark charged current decays to either
$s$\ or a $d$~quarks, but only via the mixing of the quark mass eigenstates as
parametrised by the Cabibbo-Kobayashi-Maskawa (CKM) matrix elements
$V_{ts}$\ or $V_{td}$.  
If we assume that there are only three generations and that the CKM matrix
is unitary, then the 90\%\ CL limits on these two elements 
are \cite{ref: CKM limits}
\begin{eqnarray}
0.004 \le |V_{td}| \le 0.015 \qquad {\rm and } \qquad
0.030 \le |V_{ts}| \le 0.048.
\label{eq: Vts Vtd range}
\end{eqnarray}
This predicts top quark branching fractions to 
$s$\ and $d$~quarks of less than 0.3\%.  
However, if we relax the condition of unitary and/or allow for a larger
number of quark generations, then the strict limits on $V_{ts}$\ and $V_{td}$\ 
no longer apply, and the possibility exists for large top quark decay rates
to these lighter quarks.  

There are a number of standard model extensions that predict decay modes 
not involving a transition mediated by a $W^+$\ boson 
\cite{ref: top FCNC models}.  The most obvious candidates are the
flavour-changing neutral decays such as $t\rightarrow Z^\circ c$\ or
$t\rightarrow \gamma c$.  Such models therefore result in decays
that violate both standard model predictions.
There are also models that predict decay modes that always yield a 
$b$~quark in the final state, but involve a transition mediated by
something other than the $W^+$\ boson.  A popular example of this is
the decay $t \rightarrow H^+ b$, where $H^+$\ is a charged Higgs boson.
Since the decay modes of the $H^+$\ are in principle quite different from
those of the $W^+$, this would result in a different rate of 
lepton+jet and dilepton final states
coming from the \ttbar\ system.

\subsubsection{Top Quark Branching Fraction}

The measurement of top quark branching fractions is currently limited
by the rather small number of detected events, and by the large uncertainty
in the top quark production cross section.  The most sensitive measures of
the top quark branching fraction $\mBR(t\rightarrow W^+b)$\ that 
do not depend on a knowledge of the $\sigma_{\mttbar}$\ are the 
relative rate of single to double $b$~quark tags in lepton+jet
events, and the relative rates of zero, single and 
double $b$~quark tags in dilepton events.
The relative rate of zero $b$~quark tags in lepton+jet events is not
helpful in this case as this sample is contaminated with a large fraction of 
non-\ttbar\ background.

These relative rates are sensitive to
\begin{eqnarray}
{\cal R}  \equiv {{BR(t\rightarrow W^+b)}\over{BR(t\rightarrow W^+q)}} = 
{{|V_{tb}|^2}\over{|V_{tb}|^2 + |V_{ts}|^2 + |V_{td}|^2}}.
\end{eqnarray}
The fractions of zero, single and double tagged events can be related to
${\cal R}$\ by the expressions
\begin{eqnarray}
f_0 &=& (1-{\cal R}\epsilon)^2 \nonumber \\
f_1 &=& 2{\cal R}\epsilon (1-{\cal R}\epsilon) \\
f_2 &=& ({\cal R}\epsilon )^2, \nonumber
\end{eqnarray}
where $\epsilon$\ is the $b$~tagging efficiency.  
These can be solved for ${\cal R}$\ to obtain the expressions
\begin{eqnarray}
	{\cal R} &=& {{2}\over{\epsilon (f_1/f_2 + 2)}}  \\
	{\cal R} &=& {{1}\over{\epsilon (2f_0/f_1 + 1)}}, 
\end{eqnarray}
where the first expression is applicable to both the lepton+jets and
dilepton event samples, and the second applies to the dilepton sample only.

\begin{figure}
\vbox{
\vspace*{4.5in}
\vskip+1.8in
\hskip+0.0in
\includegraphics{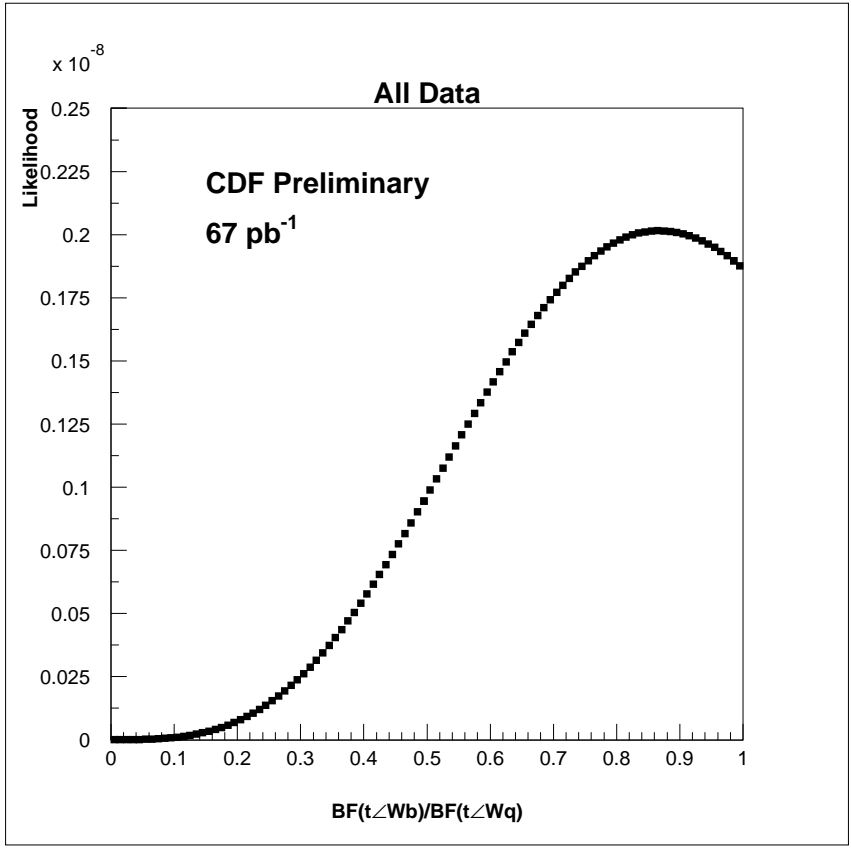}
\vskip-1.8in
}
\caption{The likelihood function of ${\cal R}$\ determined by using
the relative rate of zero, single and doubled tagged events in the CDF
dilepton data and the relative rates of single and double tagged events in
the CDF lepton+jet data.
}
\label{fig: Vtb likelihood}
\end{figure}

These relative rates of $b$~tagged events are most efficiently
combined by using a 
maximum likelihood technique to determine ${\cal R}$.  
The likelihood function that combines the CDF data from each channel is
shown in Fig.~\ref{fig: Vtb likelihood}\ as a function of ${\cal R}$.
The function peaks near unity, but has a large width that results from the
limited statistics of the sample.  From this distribution, one 
determines that 
\begin{eqnarray}
{\cal R } = 0.87 ^{+0.13}_{-0.30} ({\rm stat}) ^{+0.13}_{-0.11} ({\rm syst}),
\end{eqnarray}
where the systematic uncertainty is dominated by the uncertainty in 
$b$~tagging efficiency.  

Since ${\cal R}$\ is a ratio involving three CKM matrix elements, we
can convert this measurement into a statement about $|V_{tb}|$\ by
assuming, for example, the limits on
$V_{td}$\ and $V_{ts}$\ quoted in Eq.~\ref{eq: Vts Vtd range}.
This results in 
\begin{eqnarray}
|V_{tb}| = 0.11 ^{+0.89}_{-0.05}, 
\end{eqnarray}
which is in agreement with the standard model expectation, albeit with
large uncertainties.  The result is most directly interpreted as implying
$|V_{tb}| \gg |V_{ts}|$\ or $|V_{td}|$. 

\subsubsection{Other Aspects of Top Quark Decays}
The poor statistics of the \DZero\ and CDF samples limit the
detail with which one can study other aspects of top quark decays.
However, I would like to mention two specific studies that are
currently underway, though results are not yet available. 

The $V$-$A$\ nature of the charged current 
results in the prediction that the decay
$t\rightarrow W^+ b$\ will result in $W^+$\ bosons that are 
longitudinally polarised, that is, they will be produced with
helicity aligned transverse to their momentum vector.  This arises from
the large top quark mass, as the fraction of longitudinal polarisation is
given by
\begin{eqnarray}
{{M_{top}^2/(2M_W^2)}\over{ 1 + M_{top}^2/(2M_W^2)}}.
\end{eqnarray}
One will, with sufficient statistics,  
be able to extract this helicity information from the 
angular distribution of the charged or neutral 
lepton helicity angle measured in
the lab frame that arises from the leptonic decay of
the $W^+$\ boson.\cite{ref: top helicity}

One can also test for FCNC top decays by searching for evidence of
$Z^\circ$\ or $\gamma$\ bosons in final states such as
\begin{eqnarray}
p\bar{p} &\rightarrow& \mttbar \rightarrow Z^\circ c W^+ b \nonumber \\
p\bar{p} &\rightarrow& \mttbar \rightarrow \gamma c W^+ b \\
p\bar{p} &\rightarrow& \mttbar \rightarrow Z^\circ c Z^\circ \bar{c},
\nonumber
\end{eqnarray}
which would arise if there was an appreciable FCNC top quark
decay rate.  These final states are essentially free of
backgrounds,\cite{ref: CDF W gamma,ref: CDF Z gamma}\
so that the searches will be limited by the $Z^\circ$\ branching ratios
and the integrated luminosity.

\subsection{Top Quark Production Properties}

QCD calculations predict that top quarks should be produced with a
relatively soft \Pt\ distribution and in the central pseudorapidity
region.  Extensive theoretical studies have been done of heavy quark
production, and the theoretical uncertainties in the QCD predictions
are quite modest.  
Although there has been some theoretical concern about the number and
spectrum of additional jets arising from QCD radiation and higher-order
processes, the general consensus is that these standard model 
uncertainties do not have a large
effect on the production kinematics of top quarks.

However, there has been speculation that new physics beyond the
standard model could have an influence on the production properties of
the \ttbar\ system \cite{ref: top colour,ref: top beyond sm}. 
There are in principle
a large number of ways that such effects could be observed, which
range from deviations from QCD in the \ttbar\ production cross
section to new particle resonances that couple strongly to the
\ttbar\ system and therefore influence the kinematics of the final
state.  

\begin{figure}
\vspace*{5in}
\vbox{
\vskip+0.25in
\hskip+0.2in
\includegraphics{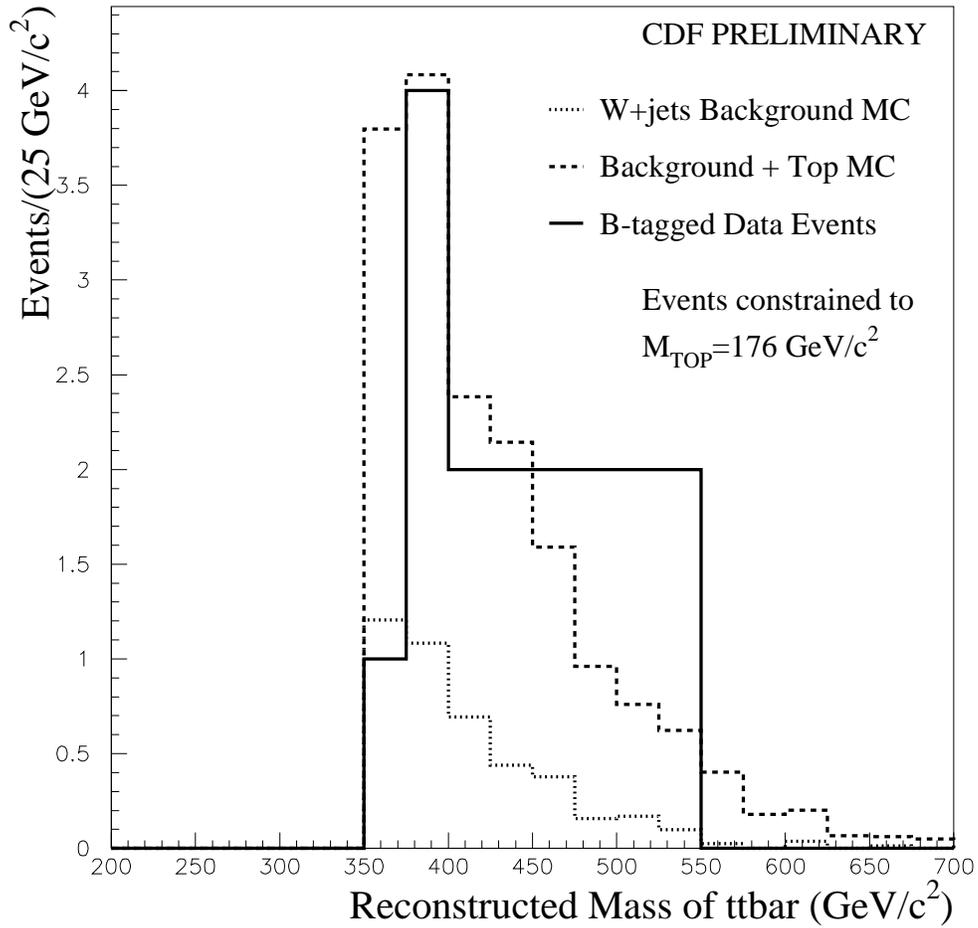}
\vskip-0.25in
\hskip-0.2in
}
\caption{The \ttbar\ invariant mass distribution of the CDF
lepton+jet sample, using the fully-reconstructed lepton+$\ge4$\ jet events.
The solid histogram is the data distribution, the heavy dashed histogram
is the standard model prediction resulting from \ttbar\ production and
the estimated background, and the light dashed histogram is the mass
distribution expected from the non-\ttbar\ background.
The top candidate events have been constrained to have a top quark 
mass equal to the CDF preliminary central value of 176~\GeVcc.
}
\label{fig: M ttbar}
\end{figure}

\begin{figure}
\vspace*{5in}
\vbox{
\vskip+0.2in
\hskip+0.2in
\includegraphics{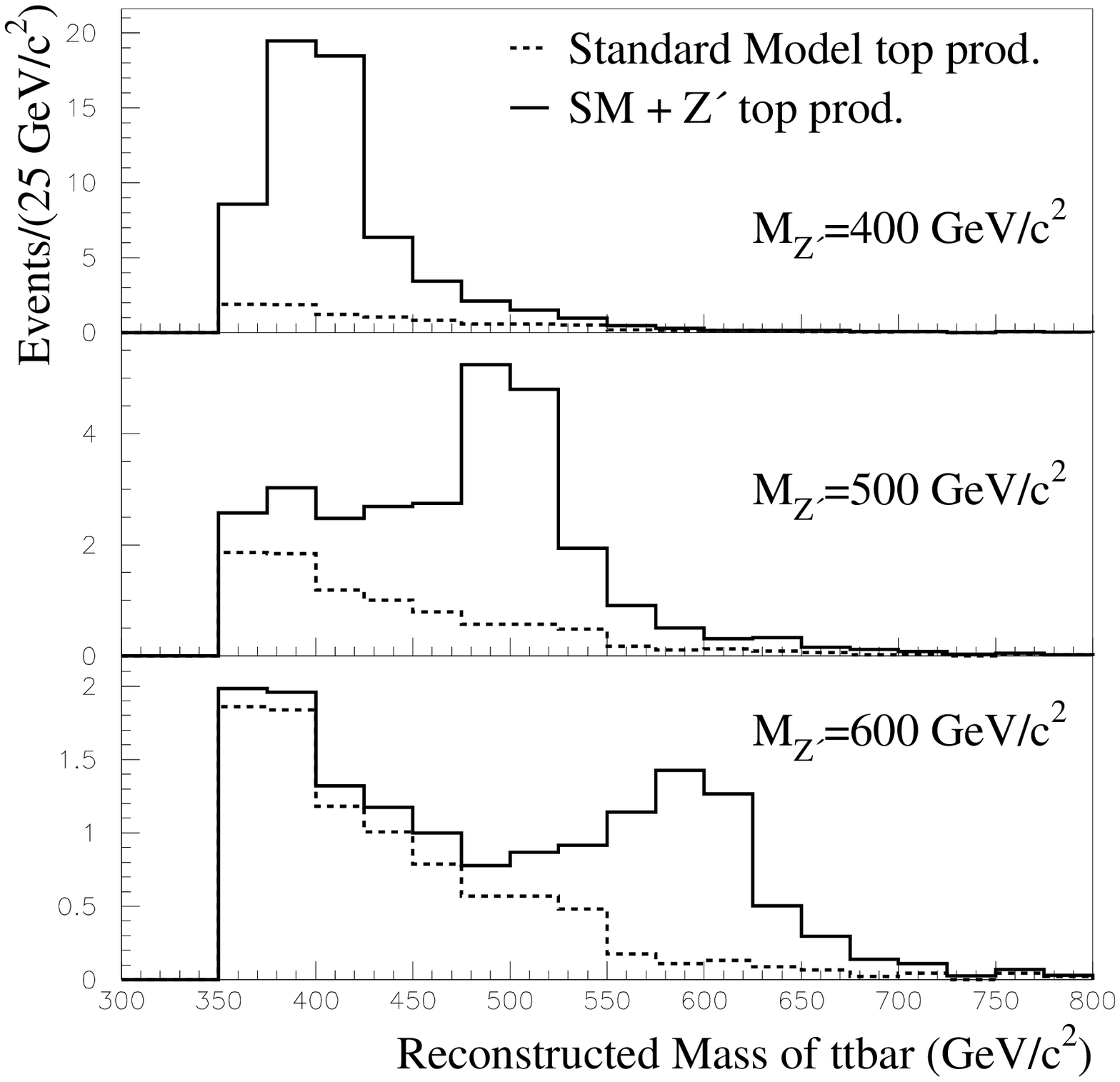}
\vskip-0.2in
\hskip-0.2in
}
\caption{The \ttbar\ invariant mass distribution that would be observed
for different $Z^\prime$\ boson masses.
The theoretical predictions include the standard model QCD prediction
combined with the $Z^\prime$\ boson coupling to the
\ttbar\ system with $Z^\prime$\ masses of 400, 500 and 600 \GeVcc.
}
\label{fig: top colour}
\end{figure}

The statistics of the CDF and \DZero\ samples limit our
ability to exclude such anomolous effects, but one study illustrates
how much we can learn from the Tevatron samples.  A resonance
coupling to the \ttbar\ system (such as a heavy neutral gauge boson,
or a $Z^\prime$) could result in an enhanced \ttbar\ production cross
section and be directly observed as an enhancement in the \ttbar\
invariant mass distribution.\cite{ref: top colour}\  
The observed \ttbar\ invariant mass distribution
from CDF is shown in Fig.~\ref{fig: M ttbar}, and is compared with
what one would expect to observe if such a $Z^\prime$\ boson does exist
in Fig.~\ref{fig: top colour}. 
Note that this phenomena is predicted to strongly enhance the total
\ttbar\ production cross section for $Z^\prime$\ boson masses of 
order 500 \GeVcc\ or less.
These data have been used to exclude at the 95\%\ CL the 
existence of a $Z^\prime$\ with mass less than $\sim470$~\GeVcc.
This limit only takes into account statistical uncertainties; 
however, it is expected to be relatively insensitive to the systematic
uncertainties that have not yet been fully characterised.

\section{Future Top Quark Studies}
\subsection{Hadron Collider Development}
Our current studies of the top quark system are based entirely on
the top quark samples that have been collected at the Fermilab
Tevatron Collider.  With approximately 100 \invpb\ of integrated
luminosity, these samples are going to remain our only direct data on
the top quark for the next three years.  

The next Tevatron Collider run, known as Run II, is scheduled to begin in
1999 and will give us at least an order of magnitude
improvement on the statistics of Run I.  This will be achieved with
the construction of the Main Injector, a new synchotron that will
replace the Tevatron's Main Ring as the accelerator and injector for
the Collider, and the construction of a new $\bar{p}$\ source.  The
Main Injector will allow significant increases in the maximum proton
density that can be accomodated during acceleration and will provide
a much larger acceptance of particles into the Tevatron Collider.  In
addition, the bunch spacing in the Tevatron Collider will be reduced
from the current 3.0 $\mu$s to 396~ns and ultimately to 132~ns.
The Tevatron maximum collision energy will also be increased by 10\%\
to 2.0~TeV by improving the capability of the cryogenic systems.

These improvements will yield an instantaneous luminosity of 
$2\times 10^{32}$\ \invcms, an order of magnitude increase from Run I
operating conditions.  Over a period of four years, the facility is
expected to provide each experiment with a data sample of 2~\invfb,
a factor of 20 increase in integrated luminosity over Run I.
The increase in centre of mass energy results in a 30\%\ increase in
the \ttbar\ yield, so an overall factor of 25 in produced top quark
pairs is therefore expected.

The next step in top quark
studies at hadron colliders will involve the Large Hadron Collider
(LHC) currently under construction at CERN and scheduled for turn-on
around 2004.  The LHC, ultimately operating at $\sqrt{s}=14$\ TeV, will
allow very high statistics studies due to the much larger \ttbar\
production cross section and the much larger luminosity.
The increased collision energy results in a \ttbar\ production
cross section of 1~nb, or a factor of 100 increase over the Run
II production cross section.  Even at relatively low initial
luminosities of $10^{32}$\ to $10^{33}$~\invcms, the LHC will be
producing top quarks at rates between 100 to 1000 times higher
than the Tevatron during Run II.  Although one has to take
care in making direct comparisons due to the significantly more
complex interactions that take place at the LHC, it is clear that
this machine will have an enormous impact on what we will learn about
the top quark.

I will briefly examine the top quark physics prospects of these two
facilities in the following sections.  A more detailed
discussion of top quark physics prospects at the 
Tevatron is available\cite{ref: TeV2000 report}.

\subsection{Tevatron Studies}

The Run II top quark studies will benefit from both the much larger
time-integrated luminosities made possible by the Main Injector
and significant improvements in both the
\DZero\ and CDF detectors.  Both collaborations are upgrading their
charged particle detection systems by replacing all their
subdetectors with new devices designed with the Run I experience in
mind and optimised for Run II operating conditions.  The \DZero\
detector will now incorporate a superconducting solenoid magnet that
will allow momentum analysis of charged particles, and both detectors
will have enhanced silicon vertex tracking detectors that provide
tracking coverage of virtually the entire luminous region. The
collaborations are making other significant improvements in lepton
identification systems, both for the detection of the high \Pt\
leptons from the decay of $W^+$\ bosons produced in \ttbar\ events
and the detection of the soft leptons from $b$~quark decay.  

\subsubsection{Top Quark Event Yields}
In order to estimate the expected number of reconstructed \ttbar\ events, 
I have used the observed CDF yields of lepton+jet and dilepton events in Run
I and taken into account the following effects:
\begin{itemize}
\item Run II will provide a factor of 20 increase in integrated
luminosity.
\item The SVX tagging efficiency will be improved by approximately a
factor of 2 due to the increase in acceptance of the SVX subdetector
to cover the entire luminous region at the interaction point.
\item The soft lepton tagging efficiencies will be improved by of
order 10\%\ by extending the technique into the pseudorapidity region 
$1\le|\eta|\le 2$.
\end{itemize}
With these assumptions, the expected yield of different categories
of events are shown in Table~\ref{tab: Run II Event Yields}.
The uncertainties on these yields are relatively large and
difficult to quantify.  Although they are based on the
observed Run I event yields, the expected improvement factors in
tagging efficiency are based on extrapolations and detector
simulations.  However, they do form a relatively concrete basis on
which to estimate the impact that the Run II data samples will have
on top physics.

I have included in this table the predicted yields of 
the $Z+\ge4$\ jet samples as well.  With the
given signal event yields, we are in a regime where the control
of systematic uncertainties arising from detector effects and
background uncertainties becomes essential to further improve the
physics measurements.  The $Z+$\ jets data provides one of the key
calibration samples as it constrains the theoretical models used to
characterise the $W+$jets background to top production.  

I will conservatively assume an integrated luminosity for Run II of
1~\invfb\ for the following discussion, although many of the results
will scale in a straight-forward manner with the assumed size of the
data sample.

\begin{table}
\begin{center}
\begin{tabular}{lcc}
\hline
Channel	&	1 \invfb	&  10 \invfb \\
\hline
Tagged $W+\ge4$\ jets	       &	600  	&   6000 \\
Double tagged $W+\ge4$\ jets	& 300	  &   3000 \\
Tagged Dilepton events 	    	& 100	  &   1000 \\[0.1in]
$Z+\ge4$\ jet events		       & 200   &   2000 \\[0.05in]
\hline
\end{tabular}
\end{center}
\caption{Projected yields of observed events for 1 and 10~\invfb\ of
integrated luminosity for both the CDF and \DZero\ experiments.
}
\label{tab: Run II Event Yields}
\end{table}

\subsubsection{Run II Top Quark Cross Section}
A more precise measurement of the top quark cross section is a good
test of our understanding of perturbative QCD calculations.  In
addition, various extensions of the standard model predict that
this cross section would be enhanced and therefore could be an 
indication of ``new'' physics.

The current uncertainties in $\sigma_{\mttbar}$\ are dominated by the
low statistics in the dilepton and lepton+jets signal samples.
For Run II, these statistical uncertainties are expected to fall to
of order 5\%\ or better.  The systematic uncertainties will therefore
limit the measurement as these are currently at the level of 
30-40\%.  However, it is possible to control most of these uncertainties
as they arise from $b$~tagging efficiencies, the background estimates
and the integrated luminosity measurements. 
For example, the $b$~tagging efficiencies can be obtained directly from
the data using the rate of single to double-tagged lepton+jet events.
I therefore expect these uncertainties to scale with the integrated
luminosity.

I believe the systematic uncertainties will be limited, in fact, by how well
we can measure the integrated luminosity in Run II.  It is not clear
that we will be able to determine this quantity to better than of order
3\%, and I would therefore argue that this sets the ``floor'' on 
the systematic uncertainties on any absolute cross section measurements.
If we expect that the other systematic uncertainies then scale with
the number of observed candidate events, this implies an overall
systematic uncertainty of $\sim7$\%.

With this assumption, the overall uncertainty in the cross section 
measurement could be of order 9\%, which is considerably less than the
current uncertainties of 15-20\%\ on the standard model predictions.

\subsubsection{Top Mass Measurement}
We can conservatively estimate how well we can measure the top quark
mass in Run II by extrapolating the uncertainties on the Run I mass
measurements using the $W+\ge4$~jet sample.

Monte Carlo calculations have shown that the statistical uncertainty
on $M_{top}$\ will scale as expected like $1/\sqrt{N}$, where $N$\
is the observed number of events in the sample.  This assumes that
the relative background rates will remain the same, a reasonable
hypothesis since they are dominated by the intrinisic physics
rates and not instrumentation effects.  One therefore can expect a
statistical uncertainty on $M_{top}$\ of $\sim2$~\GeVcc.

The control of the systematic uncertainties becomes the single most
important aspect of this measurement.  The largest source of
systematic uncertainty relates to the measurement of the jet energies
of the $b$~quarks and quarks from the $W^+$~boson hadronic decays.
Perhaps the most fundamental calibration tool is the observed
$W^+$\ signal in the dijet invariant mass distribution.  However,
independent calibrations can be performed by studying the balancing
of observed energies in $Z+1$\ jet and $\gamma+1$\ jet events.
With these studies, one can reasonably expect to reduce the systematic
uncertainties arising from jet energy scales to of order 5~GeV in the
Run I data set.  Since this calibration is driven by the size of
the $Z+$jet and $\gamma+$jet samples, one can assume that this 
uncertainty will scale statistically, resulting in a contribution to
the systematic uncertainty of 1-2~\GeVcc.

The other uncertainties that effect the current mass measurement 
together total 6-7~\GeVcc\ and should also scale statistically.
Note that the largest contributions come form the understanding of
the background shapes and the biases introduced by the tagging
techniques.  We would therefore predict that these would reduce to
of 1.5-2.0~\GeVcc\ in a 1~\invfb\ data sample.  If we combine these
together in quadrature, we arrive at a top quark mass systematic
uncertainty of approximately 2.5~\GeVcc, which is still larger
than the expected statistical uncertainty.  
Further reductions in the systematic uncertainty are 
possible by, for example, using the double-tagged samples instead of
just the single-tagged events.  These data have an intrinsically
better top quark mass resolution due to the reduced combinatorial
background, and have a much smaller background due to the requirement
of the second $b$~tag.  

\begin{figure}
\vspace*{4.5in}
\vbox{
\vskip+2.5in
\hskip-0.5in
\includegraphics{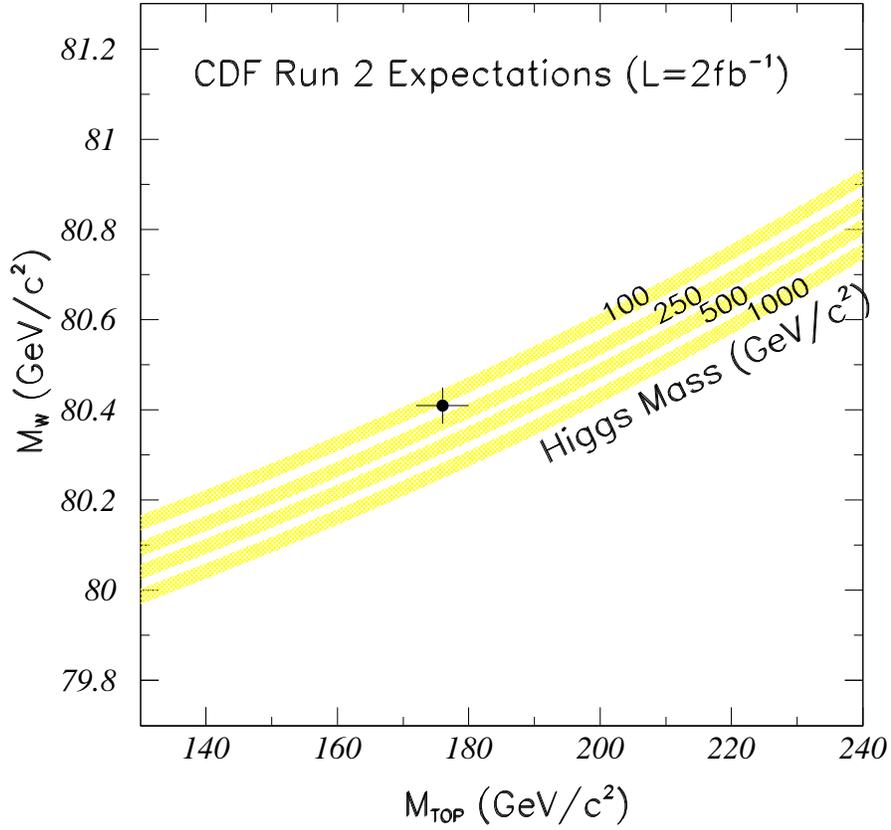}
\vskip-2.5in
\hskip+0.5in
} 
\caption{The expected precision of the top quark and $W^+$\ boson
mass measurements compared with the contours of standard model
predictions for various Higgs mass assumptions.  The central value
represents the preliminary CDF Run I top quark and $W^+$\ boson mass
measurements.
Note that the uncertainties assume an integrated luminosity of 
2~\invfb.
}
\label{fig: Mw vs Top Run II}
\end{figure}

Even without these expected improvements, the top quark mass
uncertainty will be $\sim 3$~\GeVcc, when we combine both systematic
and statistical uncertainties in quadrature.  
With the expected improvement in the $W^+$\ boson mass measurement
in Run II, we will have a very powerful test of the consistency of
the standard model.  This is illustrated in 
Fig.~\ref{fig: Mw vs Top Run II}, where we plot the expected top
quark mass versus the $W^+$~boson mass for various Higgs boson
masses.

\subsubsection{Top Quark Decays}

The top quark branching fraction for the decay $t\rightarrow W^+b$\ 
are most directly measured using the rates of tagged $b$~quarks in
both the lepton+jets and dilepton channels.  The current statistical
uncertainties on $\mBR(t\rightarrow W^+b)$\ is set by the $\pm20$\%\
uncertainty on the rate of tagged $W+$jet events.  This uncertainly
will scale as $1/\sqrt{N}$, where $N$\ is the number of tagged
events.  Thus, given the extrapolated event yields, we can expect the
statistical uncertainties on the tagging rates in the lepton+jets and
dilepton samples to fall to of order $\pm3$\%\ and $\pm4$\%,
respectively.  

The systematic uncertainties in these tagging rates are dominated by
the uncertainty in the $b$~tagging efficiency $\epsilon$.  In Run II, each
experiment will have on the order of $10^7$\ $B$~meson semileptonic
decays that will provide a high statistics sample of relatively pure
$b$~decays that can be used to study the efficiencies of the various
tagging techniques. 
With such large control samples, it is reasonable to expect that the
systematic uncertainty on $\epsilon$\ will scale with integrated luminosity. 

With these assumptions, a simple Monte Carlo calculation predicts
that one should be able to measure the branching fraction 
$\mBR(t\rightarrow W^+b)$\ with a precision of $\pm3$\%.  As noted
earlier, however, the constraint this places on $V_{tb}$\ depends on
the values that $V_{ts}$\ and $V_{td}$\ can take on.  If we assume
the same range of values as given in Eq.~\ref{eq: Vts Vtd range},
a Monte Carlo calculation combining both the
lepton+jets and dilepton tagging fractions would allow us to
constrain $V_{tb}\gessim 0.25$\ at 90\% CL.  This constraint should
also scale with luminosity so it will continue to improve
with additional data. 
Although this limit is not as stringent as that obtained if one
assumes unitarity of the CKM matrix, it is an important test of the
assumption that only 3 quark generations couple to the electroweak
force.  

With the larger Run II statistics, it will also be possible to 
make more precise measurements of the detailed structure of the $tW^+b$\ vertex.
For example,
the $V$-$A$\ nature of the current involved in the decay predicts that the decay
$t\rightarrow W^+ b$\ will result in $W^+$\ bosons that are 
longitudinally polarised.
One will be able to extract this helicity information from the 
angular distribution of the charged or neutral 
lepton helicity angle measured in
the lab frame.\cite{ref: top helicity}
Monte Carlo studies\cite{ref: TeV2000 report}\ indicate that this 
fraction can be measured to
of order 3\%\ or better.
This will make this a good test of the nature of the charged current
decay.  Any anomolous couplings are likely to become evident on the basis
of this measurement.

Searches for anomolous top quark decays will also be possible.
For example, assuming that the $\gamma W^+$\ final state is not background
limited, then a na\"\i ve calculation can be made assuming approximately 50\%\
detection efficiency for the $\gamma$\ from the decay $t\rightarrow \gamma c$\
or $t \rightarrow \gamma u$.
The efficiency for detecting the $\gamma$+jet final state relative to the 
3~jet final state resulting from the decay $t \rightarrow q\bar{q}^\prime b$\
would be $\sim0.5$.  With the expected lepton+jet event yields, we would
be sensitive to $\mBR(t \rightarrow \gamma q)$\ as small as 0.3\%.
Limits on decays mediated by $Z^\circ$\ bosons would suffer by a factor
of $\sim 5$\ due to the necessary
requirement of a dilepton decay of the $Z^\circ$\ boson.  These assume that the
final states are not background limited at this sensitivity, an assumption 
that is difficult to test with the current data samples.

\subsubsection{New Physics Searches}

The search for new physics will continue at the Tevatron Collider
during Run II, and the sensitivity of the \ttbar\ system will only
continue to improve with the increased event yields.  

\begin{figure}
\vspace*{5in}
\vbox{
\vskip+1.5in
\hskip+0.0in
\includegraphics{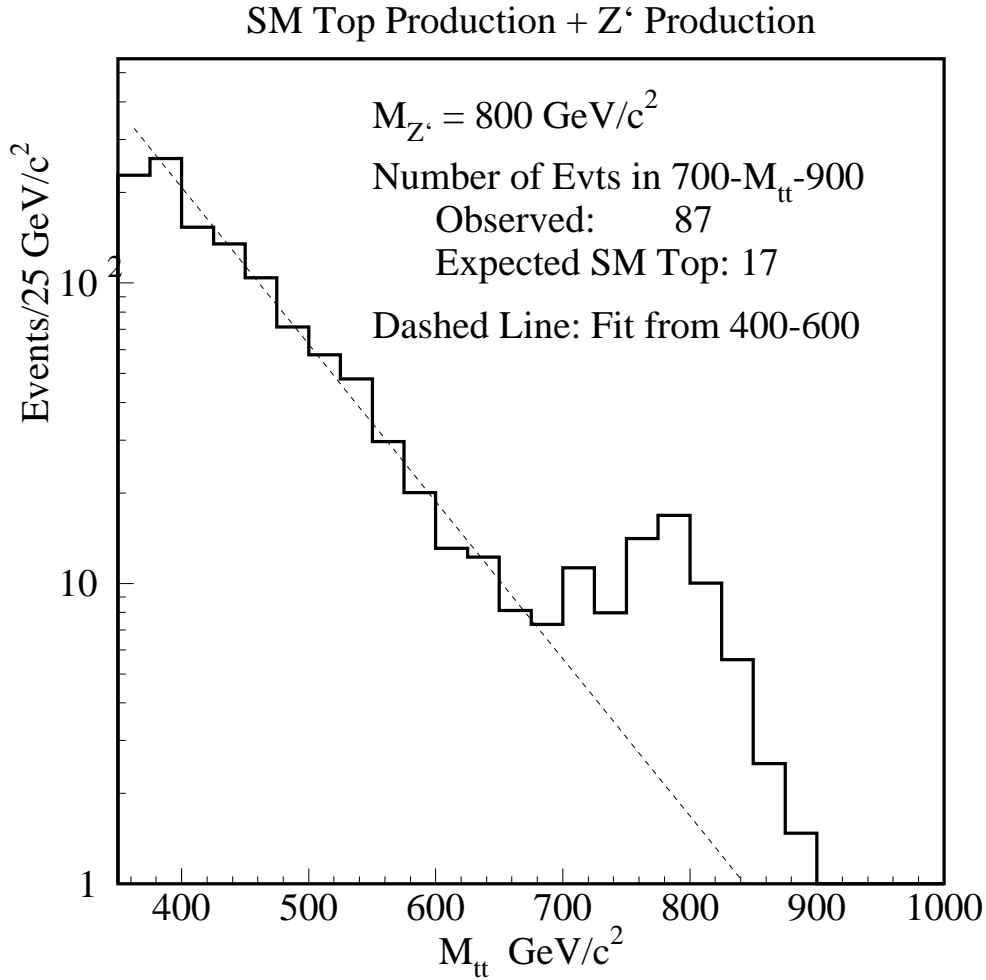}
\vskip-1.5in
\hskip+0.0in
}
\caption{The expected \ttbar\ invariant mass distribution assuming
standard model production and  the existence of a $Z^\prime$\
boson that couples to the \ttbar\ system.
}
\label{fig: top color at Run II}
\end{figure}

As one example of this, I show in Fig.~\ref{fig: top color at Run II}\
the expected \ttbar\ invariant mass distribution after 1~\invfb\ of
running, assuming the existence of a $Z^\prime$\ boson
with a mass of 800~\GeVcc.
A clear signal is visible over the standard model prediction.
One would be able to exclude the existence of such an object up to 
$Z^\prime$\ masses of order 1~\TeVcc\ during Run II.

There are other speculations about new physics that will be
addressed by studies of top production in Run II. 
The production of single top
quarks via the process $q\bar{q}^\prime \rightarrow W^\ast \rightarrow
t\bar{b}$\ is a direct way of measuring the partial width
$\Gamma(t\rightarrow W^+ b)$\ and searching for anomolous couplings
between the top quark and the electroweak bosons. 

These are only an example of the topics that will be addressed, but
they demonstrate that the Tevatron
during Run II will continue to be an exciting place to study top
quark phenomenology.

\subsection{LHC Studies}

There have been many comprehensive studies performed of the potential
for top quark physics at the much higher centre-of-mass energy
afforded by the LHC.  However, most of these studies are now dated as
they were completed prior to the discovery of top. Not only does our
current understanding of the properties of the top quark 
(most notably its mass) make many of these studies irrelevant, both
the \DZero\ and CDF collaborations have taken enormous steps
forward in understanding how to select and study \ttbar\ candidate
events in a hadron collider environment and these are not reflected
in the previous studies.

For example, the earlier SSC and 
LHC studies \cite{ref: SSC and LHC studies}\
had concluded that a precise measurement of the top quark mass
would be difficult given the large combinatorial backgrounds and
the difficulty of performing a reliable jet energy calibration. 
These studies had concluded that top quark mass measurements with a
precision of order 2-3~\GeVcc\ were possible with very large data
samples.  We now expect to acheive this level of precision at the
Tevatron with the Run II data samples.

However, I note that the LHC will produce \ttbar\
pairs at an enormous rate.  Even at a luminosity of
$10^{33}$~\invcms, the LHC will be producing of order 6000
\ttbar\ pairs per day.  Roughly speaking, an LHC experiment will be
able to collect the same number of top events in one full day of
running that would require a year's worth of data collection at the
Tevatron.  This will give an LHC experiment an enormous advantage in
statistical power over the comparable Tevatron  study.  It is 
therefore reasonable to expect that most of the studies that I
have discussed here will become very quickly systematics limited.

As a concrete example of this, the uncertainty in the top quark mass
measurement will still be dominated by the systematic uncertainties in
establishing the calorimeter energy scale.  Although the {\it in
situ}\ calibration of the calorimeter using the observed
$W\rightarrow q\bar{q}^\prime$\ invariant mass distribution
will provide a good calibration
signal, the calibration of the $b$~jet energy scales may become one
of the limiting factors.  Uncertainties arising from the
additional ``gluon'' jets in the events will also remain, though they
can be reduced by requiring, for example, two $b$~tags and only
considering lepton+4 jet events.  The ultimate precision of an LHC
mass measurement is difficult to quantify, but it is reasonable to
expect that it can be reduced to of order 1~\GeVcc\ or
perhaps less.  At this level, the top quark mass is no longer
expected to be the limiting factor in testing the consistency of the
standard model.

The very large statistics samples available at the LHC make it
possible to search for rare top quark decays.  However, such a search
will only be possible if the rare decay mode yields a sufficiently
unique signature.  For example, a signal for the rare decay
$t\rightarrow Z^\circ c$\ may ultimately be limited by the standard
model process $p\bar{p}\rightarrow W^+ Z^\circ X$\ where the
associated produced partons are $b$\ or $c$\ quark candidates.
One can expect that the sensitivity of an LHC study will be at least
an order of magnitude better than the corresponding Tevatron limit,
but this is purely speculation as a detailed study taking into
account potential backgrounds and signal efficiencies has not been
performed.

\section{Conclusions}
The hadron collider environment has proved to be quite successful in
discovering the top quark and beginning to elucidate its properties.
However, these initial Tevatron studies of the top quark are currently
statistics limited.  Both the \DZero\ and CDF collaborations have
now completed data collection for Run I and have event samples with
sensitivities of approximately 100~\invpb.  With these data, both
collaborations will be able to improve the statistical uncertainties on
the top quark cross section and mass, and they are currently involved in
additional studies that will reduce the systematic
uncertainties in these measurements.

The CDF and \DZero\ collaborations' preliminary estimates of the 
top quark mass,
$176\pm10\pm13$\ \GeVcc\ (CDF) and $199^{+19}_{-21}\pm22$\ \GeVcc\ (\DZero),
make it the heaviest known fermion in the standard model. 
The observed rate of \ttbar\ events is
consistent with standard model predictions, and make it the rarest
phenomena observed in proton-antiproton annihilations.  The very
preliminary studies of top quark production and decay properties 
have yielded results that are consistent with the standard model 
predictions.  However, additional analyses are underway and results from
the full Run I data set will yield further insights on the properties of
this unique fermion.
Because of the massiveness of this fermion, it will
be a unique probe into the physics of the standard model and what lies
beyond this theory.

The Tevatron will continue to have a monopoly on direct \ttbar\ studies for
the next eight years.  Run II, starting in 1999, will provide \ttbar\ 
samples at least 20 times larger than those available in Run I, and
will allow the first ``high statistics'' studies of the top quark.
However, the LHC will be the ultimate hadron collider for top quark studies,
as most of the standard model measurements will rapidly become systematics
limited at this machine.  In all, the future of top quark studies at
hadron colliders looks very promising indeed.

\section*{Acknowlegements}
I would like to thank the members of the \DZero\ and CDF collaborations
who kindly provided me with details of their respective analyses.
I note that the research reported here would not have
been successful without the excellent support of the staff of the
Fermilab National Accelerator Laboratory and the institutions involved
in this research.
I would also like to acknowledge the organisers of the SLAC Summer
Institute for their success in arranging such a stimulating and enjoyable
meeting.

Support for this work from the National Sciences and Engineering Research 
Council of Canada is gratefully acknowledged.

%
%


\begin{thebibliography}{10}
%
\bibitem{ref: sm}
S.~L.~Glashow, Nucl.~Phys. {\bf 22}, 579 (1961); \\
S.~Weinberg, Phys.~Rev.~Lett.~{\bf 19}, 1264 (1967); \\
A.~Salam, in {\it Elementary Particle Theory}, edited by N.~Svartholm (Almquist and
Wiksells, Stockholm, 1968), p. 367.
%
\bibitem{ref: b quark discovery}
S.~W.~Herb {\it et al.}, Phys.~Rev.~Lett. {\bf 39}, 352 (1977).
%
\bibitem{ref: LEP b quark asymmetry}
D.~Buskulic \etal\ (ALEPH Collaboration), Phys.~Lett. {\bf B335}, 99 (1994); \\
M.~Acciari \etal\ (L3 Collaboration), Phys.~Lett. {\bf B335}, 542 (1994); \\
P.~Abreu \etal\ (DELPHI Collaboration), Z.~Phys. {\bf C65}, 569 (1995); \\
G.~Alexander \etal\ (OPAL Collaboration), CERN-PPE/95-179 (1995).
%
\bibitem{ref: B->mumu limits}
R.~Ammar {\it et al.}\ (CLEO Collaboration), Phys.~Rev.~{\bf D49}, 5701 (1994).
%
\bibitem{ref: recent LEP t mass}
Review of Particle Properties, Phys.~Rev.~{\bf D50}, 1304 (1994).
%
\bibitem{ref: top searches} 
D.~Decamp {\it et al.}\ (ALEPH Collaboration), 
Phys.~Lett.~{\bf B236}, 511 (1990); \\
O.~Adriani {\it et al.}\ (L3 Collaboration), 
Phys.~Lett.~{\bf B313}, 326 (1993); \\
P.~Abreu {\it et al.}\ (DELPHI Collaboration), 
Phys.~Lett.~{\bf B242}, 536 (1990); \\
M.~Z.~Akrawy {\it et al.}\ (OPAL Collaboration), 
Phys.~Lett.~{\bf B236}, 364 (1990); \\
G.~S.~Abrams {\it et al.}\ (MARK II Collaboration), 
Phys.~Rev.~Lett.~{\bf 63}, 2447 (1989); \\
I.~Adachi {\it et al.}\ (TOPAZ Collaboration), 
Phys.~Lett.~{\bf B229}, 427 (1989); \\
S.~Eno {\it et al.}\ (AMY Collaboration), 
Phys.~Rev.~Lett.~{\bf 63}, 1910 (1989); \\
H.~Yoshida {\it et al.}\ (VENUS Collaboration), 
Phys.~Lett.~{\bf B198}, 570 (1987). 
%
\bibitem{ref: Gamma W measurements} 
F.~Abe {\it et al.}\ (CDF Collaboration), Phys.~Rev.~Lett.~{\bf 74}, 
341 (1995); \\
F.~Abe {\it et al.}\ (CDF Collaboration), Phys.~Rev.~Lett.~{\bf 73}, 
220 (1994); \\
C.~Albajar {\it et al.}\ (UA1 Collaboration), Phys.~Lett.~{\bf B253}, 503 (1991); \\
J.~Alitti {\it et al.}\ (UA2 Collaboration), Phys.~Lett.~{\bf B276}, 365 (1991).
%
\bibitem{ref: mssm}
J.~F.~Gunion, H.~E.~Haber, G.~Kane and S.~Dawson, {\it The Higgs Hunter's
Guide}\ (Addison-Wesley, New York, 1990); 
S.~L.~Glashow and E.~E.~Jenkins, Phys.~Lett. {\bf B196}, 233 (1987).
%
\bibitem{ref: charged Higgs}
F.~Abe {\it et al.}\ (CDF Collaboration), Phys.~Rev.~Lett.~{\bf 73}, 2667 (1994).
%
\bibitem{ref: Dzero limit}
S.~Abachi {\it et al.}\ (\DZero\ Collaboration), Phys.~Rev.~Lett.~{\bf 72}, 2138 (1994).
%
\bibitem{ref: CDF PRD/PRL}
F.~Abe {\it et al.}\ (CDF Collaboration), Phys.~Rev.~Lett.~{\bf 73}, 
225 (1994); \\
Phys.~Rev.~{\bf D50}, 2966 (1994).
%
\bibitem{ref: CDF/D0 Run Ib Results}
F.~Abe {\it et al.}\ (CDF Collaboration), Phys.~Rev.~Lett.~{\bf 74}, 2626 (1995);  \\
S.~Abachi {\it et al.}\ (\DZero\ Collaboration), Phys.Rev.~Lett.~{\bf 74}, 2632 (1995).
%
\bibitem{ref: single top}
D.~O.~Carlson and C.-P.~Yuan, Phys.~Lett. {\bf B306}, 386 (1993); \\
T.~Stelzer and S.~Willenbrock, ``Single Top Quark Production via 
$q\bar{q}\rightarrow t\bar{b}$,'' DTP/95/40, ILL-(TH)-95-30 (1995).
%
\bibitem{ref: Q sigmas}
P.~Nason, S.~Dawson and R.~K.~Ellis, Nucl.~Phys.~{\bf B303}, 607 (1988); \\
W.~Beenakker, H.~Kuijf and W.~L.~van~Nerven, Phys.~Rev.~{\bf D40}, 54 (1989); \\
G.~Altarelli, M.~Diemoz, G.~Martinelli and P.~Nason, Nucl.~Phys.~{\bf B308},
724 (1988).
%
\bibitem{ref: soft-gluon corrections}
E.~Laenen, J.~Smith and W.~L.~van~Neerven, Nucl.~Phys.~{\bf B369}, 
543 (1992); \\
E.~Laenen, J.~Smith and W.~L.~van~Neerven, Phys.~Lett. {\bf B321}, 254 (1994).
%
\bibitem{ref: more on soft-gluons}
E.~Berger and H.~Contopanagos, Phys.~Lett. {\bf B361}, 115 (1995);
S.~Catani, M.~L.~Mangano, P.~Nason and L.~Trentadue, ``The Top Cross Section
in Hadronic Collisions,'' CERN-TH/96-21 (1996).
%
\bibitem{ref: ttbar kinematics}
S.~Frixione, M.~Mangano, P.~Nason and G.~Ridolfi, 
Phys.~Lett. {\bf B351}, 555 (1995). 
%
\bibitem{ref: ttbar in jets}
See, for example,
P.~Azzi, invited talk at XXXIst rencontres de Moriond (QCD session), 
March 23-30, 1996, CDF/PUB/TOP/PUBLIC/3679 (1996).
%
\bibitem{ref: VECBOS calculations}
F.~A.~Berends, W.~T.~Giele, H.~Kuijf and B.~Tausk, Nucl.~Phys.~{\bf B357}, 32 (1991).
%
\bibitem{ref: Observed W+jet sigma}
F.~Abe {\it et al.}\ (CDF Collaboration), Phys.~Rev.~Lett.~{\bf 70}, 
4042 (1993). 
%
\bibitem{ref: D0 detector}  
S.~Abachi {\it et al.}\ (\DZero\ Collaboration), Nucl.~Instrum.~Methods
{\bf A350}, 73 (1994).
%
\bibitem{ref: CDF detector}
F.~Abe {\it et al.}\ (CDF Collaboration), Nucl.~Instrum.~Methods Phys.~Res.~A
{\bf 271}, 387 (1988).
%
\bibitem{ref: jet cluster algorithm}
Details of the CDF and \DZero\ jet cluster algorithms are provided
in:  \\
F.~Abe {\it et al.}\ (CDF Collaboration), Phys.~Rev.~{\bf D45}, 1448 (1992). \\
S.~Abachi {\it et al.}\ (\DZero\ Collaboration),  FERMILAB-PUB-95-020-E, July 1995,
submitted to Phys.~Rev.~{\bf D}.
%
\bibitem{ref: CDF H Analysis}
F.~Abe {\it et al.}\ (CDF Collaboration), 
FERMILAB-PUB-95-149-E, June 1995.
Submitted to Phys.~Rev.~Lett.
%
\bibitem{ref: aplanarity definition}
See, for example, V.~D.~Barger and R.~J.~N.~Phillips, ``Collider Physics,''
Addison-Wesley Publishing Co. (1987), p. 281.
%
\bibitem{ref: PDG limit calculation}
Review of Particle Properties, L.~Montanet \etal,
Phys.~Rev.~{\bf D50}, 1278 (1994).
%
\bibitem{ref: ISAJET}
F.~Paige and S.~D.~Protopopescu, BNL Report No. 38034, 1986 (unpublished).
%
\bibitem{ref: CDF 88-89 Top PRD}
F.~Abe {\it et al.}\ (CDF Collaboration), Phys.~Rev.~Lett. {\bf 68}, 447 (1992);
Phys.~Rev.~{\bf D45}, 3921 (1992).
%
\bibitem{ref: FCNC limits}
Review of Particle Properties, L.~Montanet \etal,
Phys.~Rev.~{\bf D50}, 1229 (1994).
%
\bibitem{ref: CKM limits}
Review of Particle Properties, L.~Montanet \etal,
Phys.~Rev.~{\bf D50}, 1315 (1994).
%
\bibitem{ref: top FCNC models}
H.~Fritzsch, Phys.~Lett. {\bf B224}, 423 (1989).
%
\bibitem{ref: top helicity}
G.~Kane, C.-P.~Yuan and G.~Ladinsky, Phys.~Rev.~D~{\bf 45}, 124 (1992); \\
M.~Jezabek and J.~H.~Kuhn, Phys.~Lett. {\bf B329}, 317 (1994).
%
\bibitem{ref: CDF W gamma}
F.~Abe \etal\ (CDF Collaboration), Phys.~Rev.~Lett. {\bf 74}, 1936 (1995).
%
\bibitem{ref: CDF Z gamma}
F.~Abe \etal\ (CDF Collaboration), Phys.~Rev.~Lett. {\bf 74}, 1942 (1995).
%
\bibitem{ref: top colour}
C.~T.~Hill, Phys.~Lett. {\bf B345}, 483 (1995).
%
\bibitem{ref: top beyond sm}
See, for example, 
B.~Holdom and M.~V.~Ramana, Phys.~Lett. {\bf B353}, 295 (1995).
%
\bibitem{ref: TeV2000 report}
D.~Amidei \etal, ``The TeV2000 Report: Top Physics at the Tevatron,''
CDF/DOC/TOP/PUBLIC/3265 and \DZero\ Note 2653, 1996 (unpublished).
%
\bibitem{ref: SSC and LHC studies}
See, for example, D.~Froidevaux, ``Top Quark Physics at LHC/SSC,''
CERN/PPE/93-148, published in {\it '93 Electroweak Interactions
and Unified Theories}, Ed. J. Tran Thanh Van, Editions Frontieres
(1993) 509-526; \\
W.~W.~Armstrong \etal\ (ATLAS Collaboration), ``ATLAS Technical Proposal,''
CERN/LHCC/94-43 (1994); \\
D.~Denegri \etal\ (CMS Collaboration), CERN-PPE-95-183 (1995); \\
E.~L.~Berger \etal\ (SDC Collaboration), ``SDC Technical Design Report,''
SDC-92-201 (1992).
%

\end{thebibliography}
\end{document}